\documentclass[preprint,12pt]{elsarticle}

\usepackage{amssymb}
\usepackage{amsmath}

\usepackage{float}
\usepackage{subcaption}
\usepackage{tikz}
\usetikzlibrary{calc, fit}

\usepackage{array}
\usepackage{makecell}
\usepackage{seqsplit}
\usepackage{graphicx}
\usepackage{hyperref}
\usepackage{booktabs}

\journal{Computers in Biology and Medicine}

\begin{document}

\begin{frontmatter}

\title{Limitations of Sequence-Based Protein Representations for Parkinson's Disease Classification: A Leakage-Free Benchmark}

\author[inst1]{César Jesús Núñez-Prado}

\author[inst2]{Grigori Sidorov\corref{cor1}}
\ead{sidorov@cic.ipn.mx}

\author[inst1]{Liliana Chanona-Hernández}

\cortext[cor1]{Corresponding author}

\affiliation[inst1]{
    organization={Higher School of Mechanical and Electrical Engineering, Instituto Politécnico Nacional},
    city={Mexico City},
    country={Mexico}
}

\affiliation[inst2]{
    organization={Research Center for Computing, Instituto Politécnico Nacional},
    city={Mexico City},
    country={Mexico}
}

\begin{abstract}
	
	\textbf{Background and Objective:}
	Reliable molecular biomarkers for Parkinson's disease remain difficult to identify due to the multifactorial nature of the disorder. Although protein primary sequences are a universal and readily accessible source of biological information, it remains unclear whether sequence information alone contains sufficient signal for robust disease-related classification. This study evaluates the discriminative capacity of sequence-derived protein representations for Parkinson's disease-associated protein classification.
	
	\textbf{Methods:}
	A rigorous and leakage-free benchmark was designed comparing amino acid composition, k-mers, physicochemical descriptors, hybrid representations, and protein language model embeddings. All experiments were evaluated under a nested stratified cross-validation framework to ensure unbiased performance estimation.
	
	\textbf{Results:}
	The best-performing configuration (ProtBERT + MLP) achieved an F1-score of 0.704 $\pm$ 0.028 and ROC-AUC of 0.748 $\pm$ 0.047, indicating moderate discriminative performance. Classical representations such as k-mers achieved high recall but low precision. Statistical testing showed no significant superiority among models (Friedman test, $p = 0.1749$).
	
	\textbf{Conclusions:}
	Primary sequence information alone provides limited discriminatory power for this task. Richer biological information, including structural, functional, or interaction-level descriptors, is likely required for robust Parkinson's disease modeling. This study establishes a reproducible benchmark for future research.
	
\end{abstract}


\end{frontmatter}

\section{Introduction}
\label{sec:introduction}

Parkinson's disease (PD) is a progressive neurodegenerative disorder characterized by a complex interplay of genetic, molecular, and environmental factors, which complicates its understanding, diagnosis, and treatment \cite{ref_prajjwal_2023, ref_muleiro_2024, ref_bloem_2023}. At the molecular level, PD involves multiple mechanisms, including $\alpha$-synuclein aggregation, mitochondrial dysfunction, oxidative stress, and disruptions in cellular signaling pathways \cite{ref_srinivasan_2021, ref_geibl_2024, ref_gao_2022, ref_dong_2023, ref_khan_2025, ref_blesa_2015}. This multifactorial nature suggests that disease-related signals are distributed across multiple levels of biological organization.

The identification of reliable biomarkers for PD remains an open challenge. While machine learning and deep learning approaches have shown promising results using clinical data, neuroimaging, and physiological signals \cite{ref_zarkali_2024, ref_mei_2021, diaz2024fall, ref_rabie_2025}, these methods typically rely on structured and modality-specific data that may not always be available in practical settings. In contrast, protein sequences constitute a universal and readily accessible source of biological information. However, it remains uncertain whether primary sequence information alone contains sufficient signal for robust disease-related classification.

Sequence-based modeling has been successfully applied to tasks such as protein family classification and subcellular localization \cite{bileschi2017deepfam, almagro2017deeploc}, while early approaches based on \textit{k}-mer representations enabled the extraction of local compositional patterns \cite{asgari2015protvec}. More recently, protein language models have demonstrated the ability to capture complex dependencies and contextual information from large-scale sequence data, achieving strong performance across multiple bioinformatics tasks \cite{ref_rao_2019, elnaggar2022prottrans_review, ref_rives_2021}. These advances suggest that sequence-derived representations may also be applicable to more complex disease-related classification problems.

However, despite the growing success of sequence-based methods, this assumption remains insufficiently validated in disease-oriented settings. Reported improvements may reflect model capacity, preprocessing choices, or evaluation protocols rather than genuine increases in biological discriminative signal. In particular, the absence of controlled and leakage-free comparative studies limits the ability to isolate the true contribution of sequence-derived representations from confounding factors.

To address this gap, we present a rigorous and leakage-free benchmark for the controlled evaluation of representations derived exclusively from protein primary sequences in Parkinson's disease-associated protein classification. The proposed benchmark compares classical descriptors, such as amino acid composition and \textit{k}-mers, with modern protein language model embeddings under a unified nested cross-validation framework designed to ensure unbiased performance estimation.

Rather than primarily maximizing predictive performance, our objective is to characterize the intrinsic discriminative capacity of sequence-derived representations and to identify their limitations in complex disease-related settings.

The main contributions of this work are as follows:

\begin{itemize}
	\item We introduce a leakage-free experimental framework based on nested cross-validation for the controlled evaluation of sequence-derived representations.
	
	\item We provide a reproducible benchmark for the controlled and fair comparison of classical descriptors, \textit{k}-mer-based representations, hybrid feature spaces, and protein language model embeddings within a unified protocol.
	
	\item We analyze the effect of dimensionality reduction through feature selection using a genetic algorithm, showing that reducing redundancy does not overcome the intrinsic limitations of sequence-based discrimination.
	
	\item We establish an empirical baseline that explicitly characterizes the limits of primary sequence information for Parkinson's disease-related classification tasks.
\end{itemize}

\section{Related Work}
\label{sec:relacionado}

Research on Parkinson's disease has explored a wide range of data modalities for biomarker identification, including neuroimaging, biological fluids, and clinical data. While these approaches have shown promising results, they often suffer from limited generalization, restricted data availability, and challenges in real-world applicability \cite{ref_zarkali_2024}. Machine learning and deep learning methods applied to physiological signals and clinical variables have further improved diagnostic performance \cite{ref_mei_2021, ref_rabie_2025}, but their reliance on structured and often costly data sources constrains scalability and reproducibility. In contrast, sequence-based approaches offer a more accessible alternative, although their discriminative capacity in complex disease settings remains unclear.

In bioinformatics, protein primary sequences have been widely used to construct feature representations for classification tasks. Classical descriptors, such as amino acid composition and pseudo-amino acid composition, capture global sequence properties but are inherently limited in representing higher-order dependencies \cite{ref_chou_2011}. Alignment-free methods enable efficient large-scale comparisons \cite{ref_zielezinski_2017}, yet they primarily model local or frequency-based patterns and often fail to capture the structural and functional complexity underlying biological processes.

Deep learning approaches have enabled end-to-end classification directly from sequence data. Convolutional neural networks, for instance, have demonstrated strong performance in tasks such as protein family classification and subcellular localization \cite{bileschi2017deepfam, almagro2017deeploc}. However, these approaches typically rely on task-specific architectures and training procedures, making it difficult to determine whether performance gains arise from improved representations or increased model capacity.

The introduction of distributed representations marked a transition toward more expressive sequence modeling. Methods such as ProtVec represent sequences through \textit{k}-mer embeddings inspired by natural language processing, capturing contextual relationships between residues \cite{asgari2015protvec}. More recently, protein language models such as TAPE, ProtTrans, and ESM have achieved state-of-the-art performance across multiple bioinformatics tasks by modeling long-range dependencies and complex sequence patterns \cite{ref_rao_2019, elnaggar2022prottrans_review, ref_rives_2021}. These models have also demonstrated the ability to encode structural information directly from sequence \cite{ref_lin_2023}. However, in many cases, these approaches rely on fine-tuning, auxiliary datasets, or task-specific optimization pipelines, making it difficult to isolate the contribution of the representation itself.

Despite these advances, current evaluation practices often conflate the effects of representation, model capacity, and experimental design. As a result, improvements in predictive performance cannot be attributed solely to the quality of the underlying representation, but may instead reflect confounding factors such as data preprocessing, hyperparameter optimization, or training strategies. This limitation hinders a clear understanding of the intrinsic discriminative power of sequence-derived features.

Furthermore, prior studies have shown that predicting biological function from primary sequence alone remains a challenging problem \cite{radivojac2013large}. In complex disease contexts such as Parkinson's disease, relevant signals are likely distributed across multiple biological levels, including structure, molecular interactions, and cellular context, which are not explicitly encoded in sequence data.

Overall, existing work has predominantly focused on maximizing predictive performance through increasingly complex models or multimodal integration strategies, while comparatively less attention has been given to isolating the contribution of sequence-derived representations under controlled conditions.

\begin{table*}[t]
	\centering
	\caption{Representative literature highlighting the gap between Parkinson's disease modeling studies and sequence-based protein representation research. Categories are derived from the scope explicitly indicated by the cited works.}
	\label{tab:related_work_gap}
	\scriptsize
	\setlength{\tabcolsep}{4pt}
	\renewcommand{\arraystretch}{1.15}
	\resizebox{\textwidth}{!}{%
		\begin{tabular}{p{3.0cm} p{2.8cm} p{3.0cm} p{3.8cm} p{3.2cm}}
			\hline
			\textbf{Study} & \textbf{Parkinson-focused} & \textbf{Sequence-based} & \textbf{Primary contribution} & \textbf{Gap relevance} \\
			\hline
			
			Zarkali et al. \cite{ref_zarkali_2024}
			& Yes
			& No
			& Neuroimaging and fluid biomarkers for PD
			& Disease focus without sequence benchmarking \\
			
			Mei et al. \cite{ref_mei_2021}
			& Yes
			& No
			& Machine learning for PD diagnosis
			& Disease modeling with non-sequence data \\
			
			Rabie and Akhloufi \cite{ref_rabie_2025}
			& Yes
			& No
			& ML/DL methods for PD detection
			& Disease focus with heterogeneous modalities \\
			
			Asgari and Mofrad \cite{asgari2015protvec}
			& No
			& Yes
			& Distributed biological sequence embeddings
			& Sequence modeling without PD focus \\
			
			Rao et al. \cite{ref_rao_2019}
			& No
			& Yes
			& Protein transfer learning benchmark
			& Sequence benchmarking without PD focus \\
			
			Rives et al. \cite{ref_rives_2021}
			& No
			& Yes
			& Large-scale protein language modeling
			& Advanced sequence modeling without PD focus \\
			
			\textbf{This work}
			& Yes
			& Yes
			& Leakage-free benchmark of multiple protein sequence representations for PD-associated proteins
			& Connects disease classification and sequence benchmarking \\
			
			\hline
	\end{tabular}}
\end{table*}

As summarized in Table~\ref{tab:related_work_gap}, prior studies have typically focused either on Parkinson's disease modeling using non-sequence data or on sequence-based protein representation learning without a disease-specific Parkinson setting. In contrast, the present work bridges these directions through a controlled and leakage-free benchmark centered on Parkinson-associated protein classification.

\section{Methodology}
\label{sec:metodologia}

\begin{figure*}[t]
\centering
\begin{tikzpicture}[
    x=\textwidth,
    y=1cm,
    font=\tiny,
    box/.style={
        draw,
        rounded corners,
        align=left,
        inner sep=3pt,
        text width=0.155\textwidth,
        minimum height=1.45cm
    },
    smallbox/.style={
        draw,
        rounded corners,
        align=center,
        inner sep=3pt,
        text width=0.085\textwidth,
        minimum height=0.70cm
    },
    arrow/.style={->, thick}
]

\pgfmathsetlengthmacro{\fullboxw}{0.155*\textwidth + 6pt}

\node[smallbox] (data) at (0.0425,0) {\textbf{Dataset}\\(UniProt)};

\node[box] (prep) at (0.240667,0) {
    \textbf{Preprocessing}\\[0.8mm]
    $\bullet$ Sequence validation\\
    $\bullet$ Ambiguous residue check\\
    $\bullet$ Leakage-free transformations
};

\node[box] (base) at (0.473833,0) {
    \textbf{Base representations}\\[0.8mm]
    $\bullet$ Sequence length / log-length\\
    $\bullet$ Amino acid composition (20 aa)\\
    $\bullet$ \textit{k}-mers\\
    $\bullet$ Physicochemical properties
};

\node[box] (adv) at (0.707,0) {
    \textbf{Advanced representations}\\[0.8mm]
    $\bullet$ 5 most frequent amino acids\\
    $\bullet$ Hybrid representation\\
    $\bullet$ ProtBERT embeddings
};

\draw[arrow] (data.east) -- (prep.west);
\draw[arrow] (prep.east) -- (base.west);
\draw[arrow] (base.east) -- (adv.west);

\node[box, anchor=west] (stage1) at (data.west |- 0,-3.5) {
    \textbf{Exploratory analysis}\\[0.8mm]
    $\bullet$ Length distribution\\
    $\bullet$ Amino acid composition\\
    $\bullet$ PCA-based separability analysis
};

\coordinate (stage2pos) at (adv.east |- 0,-3.5);

\node[box, anchor=east] (stage2) at (stage2pos) {
    \textbf{Evaluation on base representations}\\[0.8mm]
    $\bullet$ Evaluation using nested CV\\
    \hspace*{2mm}(see Fig.~\ref{fig:nested_cv})\\[0.8mm]
    \textbf{Supervised models selected}\\
    $\bullet$ Logistic Regression\\
    $\bullet$ SVM\\
    $\bullet$ KNN\\
    $\bullet$ Random Forest
};

\path let
    \p1 = (stage1.east),
    \p2 = (stage2.west),
    \n1 = {(\x2 - \x1 - 2*\fullboxw)/3}
in
    node[box, anchor=west] (cluster)
        at ($(stage1.east)+(\n1,0)$ |- 0,-3.5) {
        \textbf{Clustering}\\[0.8mm]
        $\bullet$ K-Means\\
        $\bullet$ Agglomerative
    }
    node[box, anchor=west] (baseline)
        at ($(stage1.east)+(2*\n1+\fullboxw,0)$ |- 0,-3.5) {
        \textbf{Baseline supervised models}\\[0.8mm]
        $\bullet$ Logistic Regression\\
        $\bullet$ SVM\\
        $\bullet$ KNN\\
        $\bullet$ Random Forest\\
        $\bullet$ MLP (shallow, intermediate, deep)
    };

\coordinate (mid12) at ($(base.south)!0.5!(stage2.north)$);
\coordinate (barL) at (stage1.center |- mid12);
\coordinate (barR) at (stage2.center |- mid12);

\draw[thick] (barL) -- (barR);

\draw[arrow] (base.south) -- (base.south |- mid12);

\draw[arrow] (barL) -- (stage1.north);
\draw[arrow] (barR) -- (stage2.north);

\draw[arrow] (cluster.north |- mid12) -- (cluster.north);
\draw[arrow] (baseline.north |- mid12) -- (baseline.north);

\node[box, text width=0.32\textwidth] (stage3) at ($(cluster |- 0,-7.0) + (-0.08,0)$) {
    \textbf{Evaluation on advanced representations}\\[0.8mm]
    $\bullet$ Evaluation using nested CV\\
    \hspace*{2mm}(see Fig.~\ref{fig:nested_cv})\\[0.8mm]
    \textbf{Supervised models selected}\\
    $\bullet$ KNN\\
    $\bullet$ SVM\\
    $\bullet$ MLP (shallow, intermediate, deep)
};

\coordinate (mid23) at ($(stage2.south)!0.5!(stage3.north)$);

\coordinate (bar23L) at ([xshift=-4mm]stage3.west |- mid23);

\coordinate (bar23R) at (stage2.center |- mid23);

\draw[thick] (bar23L) -- (bar23R);

\draw[arrow] (cluster.south) -- (cluster.south |- mid23);
\draw[arrow] (baseline.south) -- (baseline.south |- mid23);
\draw[arrow] (stage2.south) -- (stage2.south |- mid23);

\node[box, text width=0.20\textwidth] (ga) at ($(baseline |- 0,-7.0) + (0.07,0)$) {
    \textbf{\textit{k}-mer optimization}\\[0.8mm]
    Genetic algorithm on \textit{k}-mers
};

\draw[arrow] (ga.west) -- (stage3.east);

\coordinate (vtop) at ($(adv.south)!0.5!(stage2.north)$);

\coordinate (vbot) at (ga.center);

\draw[thick] ([xshift=5mm]stage2.east |- vtop) -- ([xshift=5mm]stage2.east |- vbot);

\draw[arrow] ([xshift=5mm]stage2.east |- vbot) -- (ga.east);

\coordinate (hadvL) at (adv.center |- vtop);
\coordinate (hadvR) at ([xshift=5mm]stage2.east |- vtop);

\draw[thick] (hadvL) -- (hadvR);

\draw[thick] (adv.south) -- (adv.south |- vtop);

\draw[arrow] (stage3.west) -- (stage3.west -| bar23L);

\node[box, text width=0.18\textwidth] (metrics) at (stage3 |- 0,-10.2) {
    \textbf{Metrics evaluation}\\[0.8mm]
    $\bullet$ Accuracy\\
    $\bullet$ Precision\\
    $\bullet$ Recall\\
    $\bullet$ F1-score\\
    $\bullet$ ROC-AUC\\
    $\bullet$ PR-AUC\\
    $\bullet$ Specificity
};

\node[box, text width=0.31\textwidth] (final) at (ga |- 0,-10.2) {
    \textbf{Final analysis}\\[0.8mm]
    $\bullet$ Comparative performance analysis\\
    $\bullet$ Confusion matrices\\
    $\bullet$ Error analysis\\
    $\bullet$ Friedman test
};

\draw[arrow] (metrics.east) -- (final.west);

\draw[thick] (bar23L) -- (bar23L |- metrics.center);
\draw[arrow] (bar23L |- metrics.center) -- (metrics.west);

\end{tikzpicture}
\caption{Experimental pipeline used in this work.}
\label{fig:pipeline}
\end{figure*}
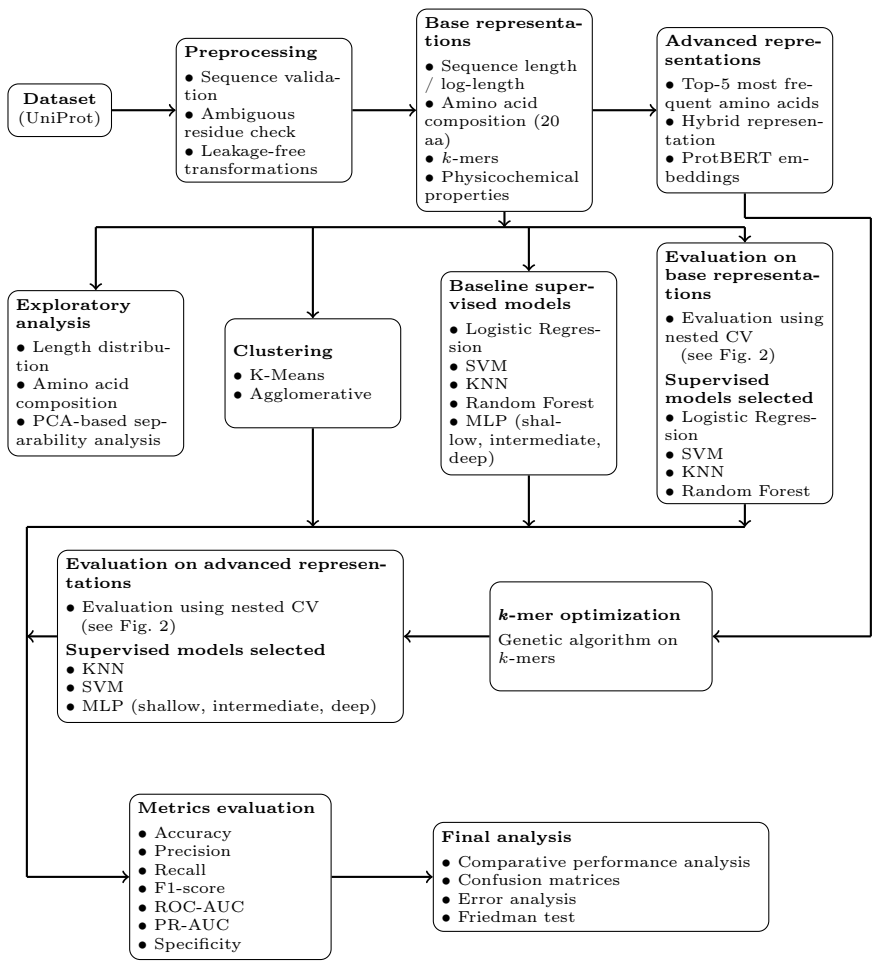

Figure~\ref{fig:pipeline} summarizes the experimental workflow, from dataset construction to model evaluation. The proposed pipeline enables a controlled and systematic comparison of multiple representations derived exclusively from protein primary sequences.

The considered representations span different levels of abstraction, including global descriptors, local compositional patterns, physicochemical properties, hybrid feature spaces, and embeddings obtained from protein language models. Additionally, a wrapper-based feature selection strategy using a genetic algorithm is applied to the \textit{k}-mer representation to analyze the impact of dimensionality reduction in high-dimensional spaces.

All representations are evaluated under a unified experimental protocol using the same set of supervised models and a stratified nested cross-validation scheme (5 outer folds for performance estimation and 3 inner folds for hyperparameter optimization). Feature selection and all data-dependent transformations are incorporated exclusively within training folds, ensuring strict separation between training and evaluation and preventing data leakage.

The entire pipeline is restricted to information derived from the primary sequence, without incorporating structural, functional, or evolutionary data. This design explicitly isolates the contribution of sequence-derived representations and enables a direct assessment of their intrinsic discriminative capacity.

\subsection{Dataset}

A curated dataset of human protein sequences was constructed to address a binary classification task between Parkinson's disease-associated proteins and control proteins.

The sequences were obtained from the UniProt (Universal Protein Resource) database, considering only proteins corresponding to human organisms (\textit{Homo sapiens}). Parkinson-associated proteins were identified using keyword-based queries related to the disease, such as ``Parkinson'' and ``Parkinson's disease'', combined with an organism filter.

From the retrieved results, only complete protein sequences were retained. During the curation process, duplicated and redundant records were removed. Additionally, a duplicated entry appearing across both classes was identified and eliminated, ensuring that no identical protein sequence is shared between classes.

The control protein set was constructed from human proteins not explicitly associated with Parkinson's disease, randomly selected from UniProt while avoiding direct overlap with the positive set. This strategy approximates a realistic discrimination scenario; however, it does not guarantee the absence of indirect associations due to incomplete or ambiguous annotations. More broadly, the assigned labels reflect curated disease associations derived from publicly available biological databases rather than experimentally validated causal mechanisms. Therefore, the task should be interpreted as a pragmatic benchmarking setting for disease-associated proteins rather than as a direct causal inference problem.

The final dataset consists of 304 proteins, evenly distributed across two classes: 152 Parkinson-associated proteins and 152 control proteins. This class balance reduces bias during model training and facilitates the interpretation of evaluation metrics. A summary of the main dataset characteristics is presented in Table~\ref{tab:dataset}.

Each record includes a unique identifier, a functional description, and the corresponding amino acid sequence represented using the standard 20-amino-acid alphabet. Protein length was also considered as a descriptive feature. The sequences exhibit high variability in length, motivating the use of representations that are independent of this property in later stages of the analysis.

Dataset integrity was verified by ensuring the absence of missing values, the validity of sequences, and consistency after duplicate removal. The data were downloaded from UniProt in March 2026.

\begin{table}[t]
	\centering
	\small
	\caption{Dataset summary}
	\label{tab:dataset}
	\begin{tabular}{l c}
		\hline
		\textbf{Feature} & \textbf{Value} \\
		\hline
		Total proteins & 304 \\
		Parkinson-associated proteins & 152 \\
		Control proteins & 152 \\
		Organism & \textit{Homo sapiens} \\
		Average length (aa) & 521 \\
		Standard deviation (aa) & 496 \\
		Minimum length (aa) & 26 \\
		Maximum length (aa) & 4678 \\
		\hline
	\end{tabular}
\end{table}

Table~\ref{tab:dataset_ejemplo} presents representative examples of records from both classes, illustrating the variability in sequence length and composition.

\begin{table*}[t]
	\centering
	\caption{Representative examples of dataset records. Only initial fragments of protein sequences are shown for illustration purposes.}
	\label{tab:dataset_ejemplo}
	
	\small
	\setlength{\tabcolsep}{4pt}
	\renewcommand{\arraystretch}{1.2}
	
	\resizebox{\textwidth}{!}{%
		\begin{tabular}{p{2.5cm} p{4.5cm} p{1.5cm} p{6.5cm}}
			
			\hline
			\textbf{ID} & \textbf{Description} & \centering\textbf{Length} & \textbf{Sequence fragment} \tabularnewline
			\hline
			
			A7MD48 (SRRM4) & Serine/arginine repetitive matrix protein 4 & \centering 611 &
			\makecell[l]{MASVQQGEKQLFEKFWRTGTFKA\\VATPRP} \tabularnewline
			
			O00217 (NDUFS8) & NADH dehydrogenase iron-sulfur protein 8 & \centering 210 &
			\makecell[l]{MRCLTTPMLLRALAQAARAGPGG\\RSLHSS} \tabularnewline
			
			O00592 (PODXL) & Podocalyxin & \centering 558 &
			\makecell[l]{MRCALALSALLLLLSTPPLLPSP\\SPSPSP} \tabularnewline
			
			P49279 (NRAMP1) & \makecell[l]{Natural resistance-associated\\ macrophage protein 1} & \centering 550 &
			\makecell[l]{MITGDKGPQRLSGSSYGSISSPT\\SPSPGQ} \tabularnewline
			
			Q60568 (PLOD3) & \makecell[l]{Multifunctional procollagen\\ lysine hydroxylase 3} & \centering 738 &
			\makecell[l]{MTSSGPGPRFLLLLLLPLLPPAA\\SASDRP} \tabularnewline
			
			Q96L34 (MARK4) & MAP/microtubule affinity-regulating kinase 4 & \centering 752 &
			\makecell[l]{MSSRTVLAPGNDRNSDTHGTLGS\\GRSSDK} \tabularnewline
			
			\hline
		\end{tabular}%
	}
\end{table*}

The dataset used in this study is publicly available on Zenodo under a Creative Commons Attribution 4.0 (CC BY 4.0) license, ensuring full reproducibility of the experiments, and can be accessed at \url{https://doi.org/10.5281/zenodo.19327790}. In this work, we specifically used version 2.0 of the dataset, which contains 304 protein sequences after the removal of duplicated entries across classes and the identification of a sequence containing a non-standard amino acid (selenocysteine, U).

\subsection{Preprocessing and Data Integrity}

A preprocessing stage was performed to ensure data quality, consistency, and suitability for feature extraction.

First, the absence of missing values was verified across all records. Sequence integrity was then assessed by validating that all amino acid symbols belong to the standard 20-amino-acid alphabet. Sequences containing ambiguous or non-standard symbols were explicitly identified; only one such case was detected, indicating high overall data quality.

Given the high variability in protein length, a natural logarithmic transformation of the form $\log(x + 1)$ was applied. This transformation reduces distribution skewness, mitigates the influence of extreme values, and improves numerical stability when incorporating length as a feature in scale-sensitive models.

Preprocessing was intentionally restricted to essential validation and normalization steps. No external information, imputation procedures, or complex transformations were introduced. This design ensures that all subsequent analyses rely exclusively on information derived from the primary sequence and prevents the introduction of external biases.

\subsection{Experimental Design and Leakage Control}
\label{sec:exp_design}

The experimental design was defined to ensure a fair, reproducible, and leakage-free evaluation of all representations and models under strict separation between training and evaluation data.

All experiments were conducted using a stratified 5-fold cross-validation scheme, preserving the class distribution (152 Parkinson-associated proteins and 152 control proteins) across folds. Given the limited dataset size (304 proteins), stratification reduces variability in performance estimates and ensures consistent evaluation conditions.

For optimized models, a nested cross-validation strategy was employed. A 5-fold outer loop was used for performance estimation, while a 3-fold inner loop was used for hyperparameter tuning. All model selection procedures were performed exclusively within the training portion of each outer fold.

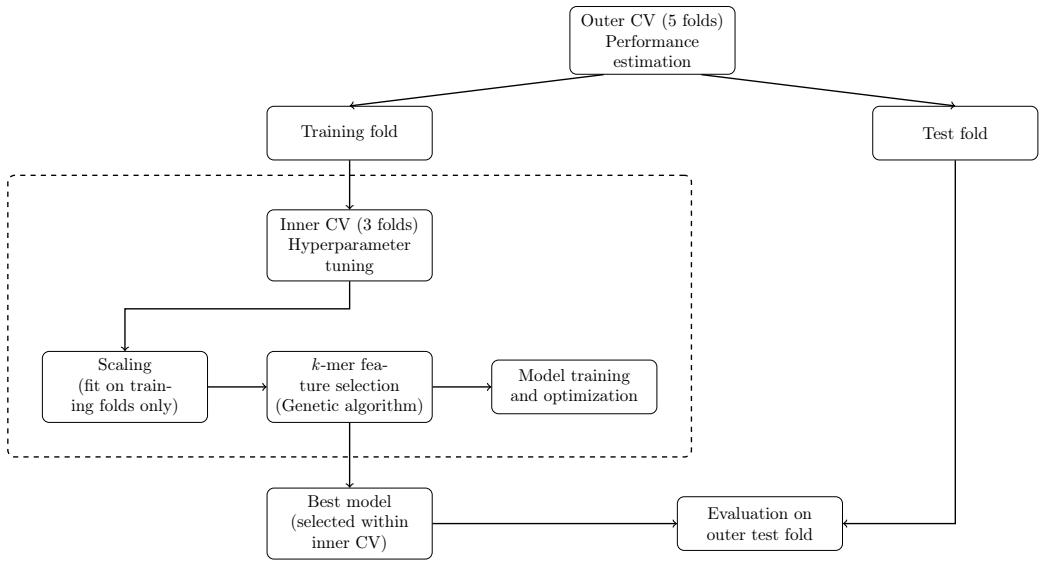
\begin{figure}[H]
	\centering
	\resizebox{\linewidth}{!}{%
	\begin{tikzpicture}[
		font=\footnotesize,
		box/.style={
			draw,
			rounded corners,
			align=center,
			text width=3.1cm,
			minimum height=1.1cm,
			inner sep=4pt
		},
		arrow/.style={->, thick},
		group/.style={
			draw,
			dashed,
			rounded corners,
			inner sep=0.7cm,
			line width=0.8pt
		}
		]
		
		\node[box] (outer) at (0,4.6) {Outer CV (5 folds)\\Performance estimation};
		
		\node[box] (train) at (-6.2,2.7) {Training fold};
		\node[box] (test)  at ( 6.2,2.7) {Test fold};

		\draw[arrow] ($(outer.south)+(-1.0,0)$) -- (train.north);
		\draw[arrow] ($(outer.south)+( 1.0,0)$) -- (test.north);
		
		\node[box] (inner) at (-6.2,0.4) {Inner CV (3 folds)\\Hyperparameter tuning};
		\draw[arrow] (train.south) -- (inner.north);
		
		\node[box] (scaling) at (-10.8,-2.5) {Scaling\\(fit on training folds only)};
		\node[box] (feature) at (-6.2,-2.5) {\textit{k}-mer feature selection\\(Genetic algorithm)};
		\node[box] (model)   at (-1.6,-2.5) {Model training\\and optimization};
		
		\draw[arrow] (inner.south) -- (-6.2,-0.9) -| (scaling.north);
		\draw[arrow] (scaling.east) -- (feature.west);
		\draw[arrow] (feature.east) -- (model.west);
		
		\node[group, fit=(inner)(scaling)(feature)(model)] {};
		
		\node[box] (best) at (-6.2,-5.3) {Best model\\(selected within inner CV)};
		\draw[arrow] (feature.south) -- (best.north);
		
		\node[box] (eval) at (2.2,-5.3) {Evaluation on\\outer test fold};
		\draw[arrow] (best.east) -- (eval.west);
		\draw[arrow] (test.south) |- (eval.east);
		
	\end{tikzpicture}%
	}
	\caption{Nested cross-validation scheme. The outer loop is used for performance estimation, while the inner loop performs hyperparameter tuning and model selection. All data-dependent transformations, including preprocessing and feature selection, are fitted exclusively on the training data within each fold, ensuring strict separation between training and evaluation and preventing data leakage.}
	\label{fig:nested_cv}
\end{figure}

Figure~\ref{fig:nested_cv} illustrates this scheme. For each outer split, the training fold is further divided into three inner folds, where hyperparameter optimization and model selection are performed. The selected model is then evaluated on the corresponding outer test fold.

All transformations that depend on the data distribution were applied exclusively within the training data of each fold. In particular, scaling procedures were fitted only on the training folds and subsequently applied to validation and test partitions, preventing information leakage.

For high-dimensional representations, additional transformations were incorporated within the inner loop. In the case of the \textit{k}-mer representation (400-dimensional dipeptide space), a wrapper-based genetic algorithm was applied for feature selection, resulting in an average of approximately 84 selected features across folds. This process was executed independently within each training fold, leading to potentially different feature subsets across folds and reflecting the dependence of the selected features on the specific training partition.

Not all representations require feature selection. While reduced representations such as the five most frequent amino acid subset are constructed directly from sequence statistics, the genetic algorithm is applied exclusively to the \textit{k}-mer representation.

Within the inner loop, the training pipeline follows a fixed sequence: scaling, feature transformation (when applicable), and model training with hyperparameter optimization. This ordering ensures that all operations are confined to the training data and prevents leakage from validation or test partitions.

All models and representations were evaluated under this protocol, enabling a consistent comparison of their discriminative capacity. The reported performance therefore reflects generalization ability without optimistic bias introduced by improper validation procedures.

\subsection{Feature Representations}

Protein sequences were converted into numerical representations to enable a direct comparison of different feature spaces under the same experimental conditions. The considered representations capture complementary aspects of the primary sequence, ranging from global descriptors to local compositional patterns and contextual embeddings.

As a first representation, protein length was incorporated as a low-dimensional global descriptor using two variables: the original sequence length and its logarithmic transformation, $\log(x+1)$. This descriptor evaluates whether coarse size-related information provides discriminative signal.

A second representation was based on amino acid composition. For each sequence, the relative frequencies of the 20 standard amino acids were computed, resulting in a 20-dimensional vector that summarizes the global residue distribution.

To capture local sequence patterns, a \textit{k}-mer representation with $k=2$ was constructed. Each sequence was represented by the relative frequencies of all possible dipeptides, yielding a 400-dimensional vector ($20^2 = 400$). This approach models short-range dependencies between adjacent residues while remaining computationally tractable.

Another representation was defined using physicochemical properties derived from residue group proportions. The following groups were considered: positively charged (K, R, H), negatively charged (D, E), polar uncharged (S, T, N, Q, C, Y), hydrophobic (A, V, I, L, M, F, W, P, G), small (A, G, S, C, T, P, D, N, V), aromatic (F, W, Y), aliphatic (A, I, L, V), sulfur-containing (C, M), and amides (N, Q). In addition, an approximate net-charge descriptor was included, defined as the difference between the proportions of positively and negatively charged residues. These variables produce a 10-dimensional feature space.

Based on these descriptors, a hybrid representation was constructed by concatenating length-based variables, amino acid composition, \textit{k}-mer frequencies, and physicochemical properties, resulting in a 432-dimensional feature space.

In addition to these predefined representations, a reduced representation based on the five most frequent amino acids per sequence was evaluated. For each protein, the five amino acids with the highest relative frequency were selected, and their corresponding frequencies were used as features, providing a compact summary of dominant residue composition.

To analyze redundancy in the high-dimensional \textit{k}-mer space, a wrapper-based feature selection strategy using a genetic algorithm was applied to the 400-dimensional dipeptide representation. Each individual was encoded as a binary vector indicating the inclusion or exclusion of each \textit{k}-mer. The fitness function was defined based on classification performance computed exclusively on training data within each fold. The evolutionary process incorporated tournament selection, one-point crossover, and bit-flip mutation.

As a result, a reduced \textit{k}-mer representation was obtained, with feature subsets selected independently within each fold. The number of selected features varied across folds, with an average of 84 features per fold, reflecting the dependence of the selected subset on the specific training partition.

An analysis of selection frequency showed that certain \textit{k}-mers were consistently selected across folds, including dipeptides such as \texttt{NC}, \texttt{HQ}, \texttt{CC}, \texttt{GA}, \texttt{YG}, and \texttt{QT}. This suggests that specific local residue combinations may contribute to the discriminative signal, although no single stable subset was consistently selected across all folds.

Finally, contextual embeddings were extracted using the pretrained ProtBERT model (\texttt{Rostlab/prot\_bert}). Each protein sequence was tokenized using the corresponding tokenizer, and a fixed-length representation of 1024 dimensions was obtained through mean pooling over the last hidden layer, considering only valid tokens and excluding padding. The model was used in inference mode without fine-tuning, enabling direct comparison with handcrafted descriptors. This design choice was intentional, as the objective of this study was not to maximize predictive performance through task-specific adaptation, but to isolate the intrinsic discriminative capacity of sequence-derived representations under a controlled and comparable evaluation framework.

\begin{table}[t]
	\centering
	\small
	\caption{Feature representations considered in this study.}
	\label{tab:dimensiones}
	\begin{tabular}{l c}
		\toprule
		\textbf{Representation} & \textbf{Dimension} \\
		\midrule
		\multicolumn{2}{c}{\textit{Classical descriptors}} \\
		Length (original + log) & 2 \\
		Amino acid composition & 20 \\
		5 most frequent amino acids (per sequence) & 5 \\
		\midrule
		\multicolumn{2}{c}{\textit{Engineered representations}} \\
		\textit{k}-mers ($k=2$) & 400 \\
		Physicochemical properties & 10 \\
		Hybrid representation & 432 \\
		\textit{k}-mers with genetic algorithm selection  & 84* \\
		\midrule
		\multicolumn{2}{c}{\textit{Deep contextual representations}} \\
		ProtBERT embeddings & 1024 \\
		\bottomrule
		\multicolumn{2}{l}{\footnotesize *Average number of selected features across folds.}
	\end{tabular}
\end{table}

\subsection{Supervised Models}

A set of supervised classification models was evaluated to analyze the behavior of different feature representations under distinct learning paradigms. The considered models include Logistic Regression, Support Vector Machines (SVM), Random Forest, K-Nearest Neighbors (KNN), and Multilayer Perceptron neural networks (MLP).

This selection covers a range of inductive biases. Logistic Regression represents linear models, SVM provides margin-based classification with the ability to model nonlinear decision boundaries through kernel functions, KNN captures local similarity structures in the feature space, Random Forest models nonlinear interactions through ensemble learning, and MLP introduces neural architectures capable of learning complex nonlinear mappings.

For the MLP model, three architectures with increasing representational capacity were considered: a shallow configuration with a single hidden layer, an intermediate architecture with two hidden layers, and a deeper configuration with three hidden layers.

Models sensitive to feature scale, such as Logistic Regression, SVM, and KNN, were combined with scaling procedures applied exclusively within the training data of each fold, as described in Section~\ref{sec:exp_design}.

Two evaluation settings were considered. First, baseline models were evaluated using default hyperparameter configurations to establish reference performance levels. Second, optimized models were obtained through hyperparameter tuning within the inner cross-validation loop using grid search over predefined search spaces for each model, with the best configurations selected based on the F1-score.

The hyperparameter search included variations in key model parameters, such as the number of neighbors for KNN, regularization strength for Logistic Regression, kernel parameters for SVM, and ensemble size and depth for Random Forest. The optimal configurations varied across folds, reflecting the dependence of model performance on both the representation and the training partition.

For clarity, the comparative results presented in Section~\ref{sec:resultados} (Tables~\ref{tab:base_models}, \ref{tab:nested_models}, and \ref{tab:advanced_models}) report only the best-performing configurations for each representation, while all evaluated models were considered during the experimental process.

\subsubsection{Baseline Model Configuration}

Baseline models were evaluated using standard configurations without hyperparameter tuning. The goal of this stage was to establish reference performance levels across different feature representations.

All models were implemented using the \textit{scikit-learn} library and, when required, were combined with feature standardization using \texttt{StandardScaler}. A fixed random seed ($\texttt{random\_state}=42$) was used to ensure reproducibility across all experiments.

The configurations used are summarized as follows:

\begin{itemize}
	\item \textbf{Logistic Regression:} L2 regularization with default strength ($C=1.0$), optimized using the \textit{lbfgs} solver, with a maximum of 1000 iterations.
	
	\item \textbf{Support Vector Machine (RBF):} radial basis function kernel with default parameters ($C=1.0$, $\gamma=\text{scale}$), probability estimates enabled, and feature standardization applied.
	
	\item \textbf{K-Nearest Neighbors (KNN):} default configuration with $k=5$, Euclidean distance, and uniform weighting, combined with feature standardization.
	
	\item \textbf{Random Forest:} ensemble of decision trees using the default \textit{scikit-learn} configuration and a fixed random seed.
	
	\item \textbf{Multilayer Perceptron (MLP):} three architectures with increasing capacity were considered:
	\begin{itemize}
		\item \textit{Shallow:} one hidden layer with 50 neurons.
		\item \textit{Intermediate:} two hidden layers with 100 and 50 neurons.
		\item \textit{Deep:} three hidden layers with 128, 64, and 32 neurons.
	\end{itemize}
	All MLP models were trained for up to 1000 iterations using default optimization settings and standardized inputs.
\end{itemize}

\subsubsection{Hyperparameter Optimization}

Hyperparameter optimization was performed using a nested cross-validation scheme to ensure an unbiased evaluation of model performance. For each outer fold, an inner cross-validation loop was used to identify the best-performing configuration for each model based on the F1-score.

The optimization process was conducted independently for each representation and outer fold. As a result, the selected hyperparameters varied across folds, reflecting the sensitivity of model performance to the specific data partition. This variability indicates the absence of a single stable configuration and highlights the importance of evaluating models under multiple training conditions.

A predefined search space was considered for each model, including variations in key parameters such as the number of neighbors for KNN, regularization strength for Logistic Regression, kernel parameters for SVM, and structural configurations for Random Forest. The best configuration within each inner loop was selected and subsequently evaluated on the corresponding outer test fold.

Table~\ref{tab:hyperparams} summarizes the range of hyperparameter values selected across folds for the evaluated models.

\begin{table}[t]
	\centering
	\caption{Summary of optimal hyperparameters identified during nested cross-validation.}
	\label{tab:hyperparams}
	\small
	\begin{tabular}{l l}
		\toprule
		\textbf{Model} & \textbf{Selected hyperparameters (range observed)} \\
		\midrule
		KNN & $k \in \{3,5,7,9\}$, weights $\in \{\text{uniform}, \text{distance}\}$ \\
		Logistic Regression & $C \in \{0.1,1,10\}$, class\_weight $\in \{\text{balanced}, \text{None}\}$ \\
		SVM (RBF) & $C \in \{0.1,1,10\}$, $\gamma \in \{\text{scale}, 0.1, 0.01\}$ \\
		Random Forest & $n\_estimators \in \{100,200\}$, max\_depth $\in \{\text{None}, 10\}$,\\
		& min\_samples\_leaf $\in \{1,2\}$ \\
		\bottomrule
	\end{tabular}
\end{table}

Overall, hyperparameter optimization leads to modest improvements in performance, with no consistent configuration dominating across all folds and representations. This behavior suggests that model performance is influenced not only by parameter selection but also by the intrinsic characteristics of the feature space.

\subsection{Evaluation Strategy}

Model performance was assessed using metrics computed across cross-validation folds, ensuring a consistent and comparable evaluation across models and feature representations.

For baseline experiments, a stratified 5-fold cross-validation scheme was used. For optimized models, performance was estimated using the nested cross-validation strategy described in Section~\ref{sec:exp_design}, providing robust and unbiased estimates of generalization performance.

For each combination of model and representation, performance metrics were averaged across folds, and their corresponding standard deviations were computed to capture both central tendency and variability.

The evaluated metrics include \textit{accuracy}, \textit{precision}, \textit{recall}, \textit{F1-score}, ROC-AUC, and PR-AUC. In addition, sensitivity and specificity were computed to provide a more detailed characterization of class-wise performance. While threshold-based metrics quantify classification outcomes, ROC-AUC provides a threshold-independent measure of separability, and PR-AUC captures the trade-off between precision and recall.

Confusion matrices were also analyzed, both in absolute and normalized form, to examine class-level behavior and to identify systematic patterns of misclassification, such as false positives and false negatives.

\subsection{Unsupervised Analysis}
\label{sec:unsupervised}

In addition to supervised classification, an unsupervised analysis was conducted to examine whether representations derived from protein primary sequences induce an intrinsic structure aligned with the underlying class labels.

Clustering techniques were applied using K-Means and Agglomerative clustering. K-Means was used to partition the data into two clusters, corresponding to the number of classes, while Agglomerative clustering was applied using a hierarchical approach with Euclidean distance and standard linkage criteria.

To evaluate clustering quality, both internal and external validation metrics were considered. The Silhouette coefficient was used to assess cluster compactness and separation, while Adjusted Rand Index (ARI) and Normalized Mutual Information (NMI) were computed to quantify the agreement between clustering assignments and ground-truth labels.

Additionally, cluster labels were aligned with class labels to compute classification-oriented metrics, including accuracy, precision, recall, F1-score, and specificity. This allows a direct comparison between unsupervised and supervised approaches under a unified evaluation framework.

\subsection{Methodological Considerations and Scope of the Study}

This study explicitly focuses on evaluating the discriminative capacity of representations derived exclusively from protein primary sequences under controlled and consistent experimental conditions.

Protein language models, such as ProtBERT, are used in their pretrained form without task-specific fine-tuning, enabling a direct comparison with handcrafted descriptors without introducing additional sources of variability related to model adaptation.

The analysis is strictly restricted to sequence-derived features, excluding structural, functional, and evolutionary information. This constraint defines the scope of the study and allows the contribution of sequence-based representations to be isolated.

Emphasis is placed on methodological consistency through standardized validation procedures and strict separation between training and evaluation data. This design ensures that observed performance differences can be attributed to the representations themselves rather than to variations in model configuration or experimental setup.

\section{Results}
\label{sec:resultados}

This section presents the experimental results obtained under the evaluation protocol described in Section~\ref{sec:exp_design}. The analysis focuses on assessing the behavior of different sequence-based representations from complementary perspectives, including statistical characterization, geometric structure, and predictive performance.

The results are organized as follows. First, exploratory analyses describe the statistical and compositional properties of the dataset. Next, the structure of the feature space is examined through dimensionality reduction and clustering, providing insight into the intrinsic organization of the data. Finally, supervised classification results are presented, including performance comparisons across representations and models, followed by a detailed analysis of classification behavior and errors.

\subsection{Exploratory Data Analysis}

Exploratory data analysis was conducted to characterize the statistical and compositional properties of the dataset prior to model evaluation. The objective of this stage is to describe general patterns, variability, and potential overlap between classes, without introducing assumptions about predictive performance.

The distribution of protein length exhibits substantial variability in both classes, with pronounced overlap in central tendencies and dispersion. Although Parkinson-associated proteins show a higher mean length, the wide spread and long-tailed behavior limit the discriminative value of this feature when considered in isolation. This pattern is consistently observed across both boxplot and histogram representations (Figures~\ref{fig:boxplot_longitud} and~\ref{fig:hist_longitud}).

Regarding amino acid composition, class-wise averages reveal small but consistent differences across selected residues. However, these variations remain subtle and do not indicate clear class separation. The compositional profiles suggest that both classes share a similar global structure, with only minor shifts in residue frequencies (Figure~\ref{fig:aa_composition}).

Overall, the exploratory analysis indicates the absence of strong separability based on basic sequence-derived features. These observations, further supported by the statistical analysis presented in Table~\ref{tab:longitud_clase} and the corresponding visualizations, suggest that any discriminative signal is weak and likely associated with higher-order patterns rather than simple global descriptors.

\subsubsection{Protein Length Analysis}

Protein length was analyzed as a global descriptor to assess differences in scale and variability between classes. Table~\ref{tab:longitud_clase} summarizes the main descriptive statistics.

Proteins associated with Parkinson’s disease exhibit a higher average length (603.178 vs 441.007 amino acids) and greater variability (standard deviation of 662.172 vs 214.780). This increased dispersion is reflected in the extended upper range of the Parkinson class, which includes sequences up to 4678 amino acids, compared to a maximum of 956 in the control group.

\begin{table}[H]
	\centering
	\caption{Protein length statistics by class.}
	\label{tab:longitud_clase}
	\small
	\begin{tabular}{lcc}
		\toprule
		\textbf{Statistic} & \textbf{Parkinson} & \textbf{Control} \\
		\midrule
		Mean & 603.178 & 441.007 \\
		Std & 662.172 & 214.780 \\
		Min & 26.000 & 111.000 \\
		Max & 4678.000 & 956.000 \\
		\bottomrule
	\end{tabular}
\end{table}

These results indicate a more heterogeneous distribution of sequence lengths in Parkinson-associated proteins, including extreme-length cases not observed in the control group. However, despite these differences, substantial overlap between distributions limits the discriminative value of this feature.

\begin{figure}[H]
	\centering
	\includegraphics[width=0.5\linewidth]{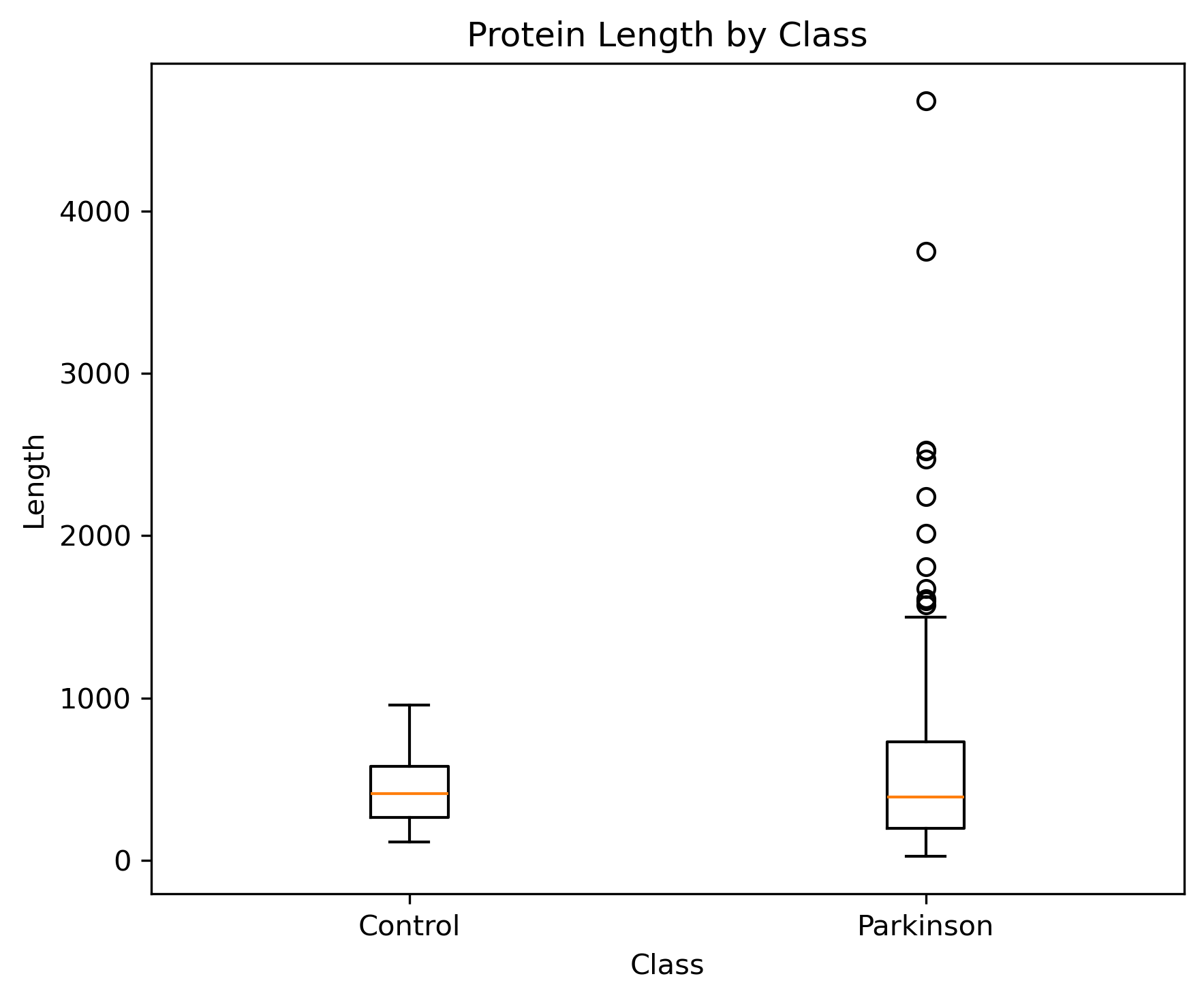}
	\caption{Distribution of protein length by class.}
	\label{fig:boxplot_longitud}
\end{figure}

Figure~\ref{fig:boxplot_longitud} shows the distribution of protein length by class, highlighting the overlap in central tendencies and dispersion.

\begin{figure}[H]
	\centering
	\includegraphics[width=0.6\linewidth]{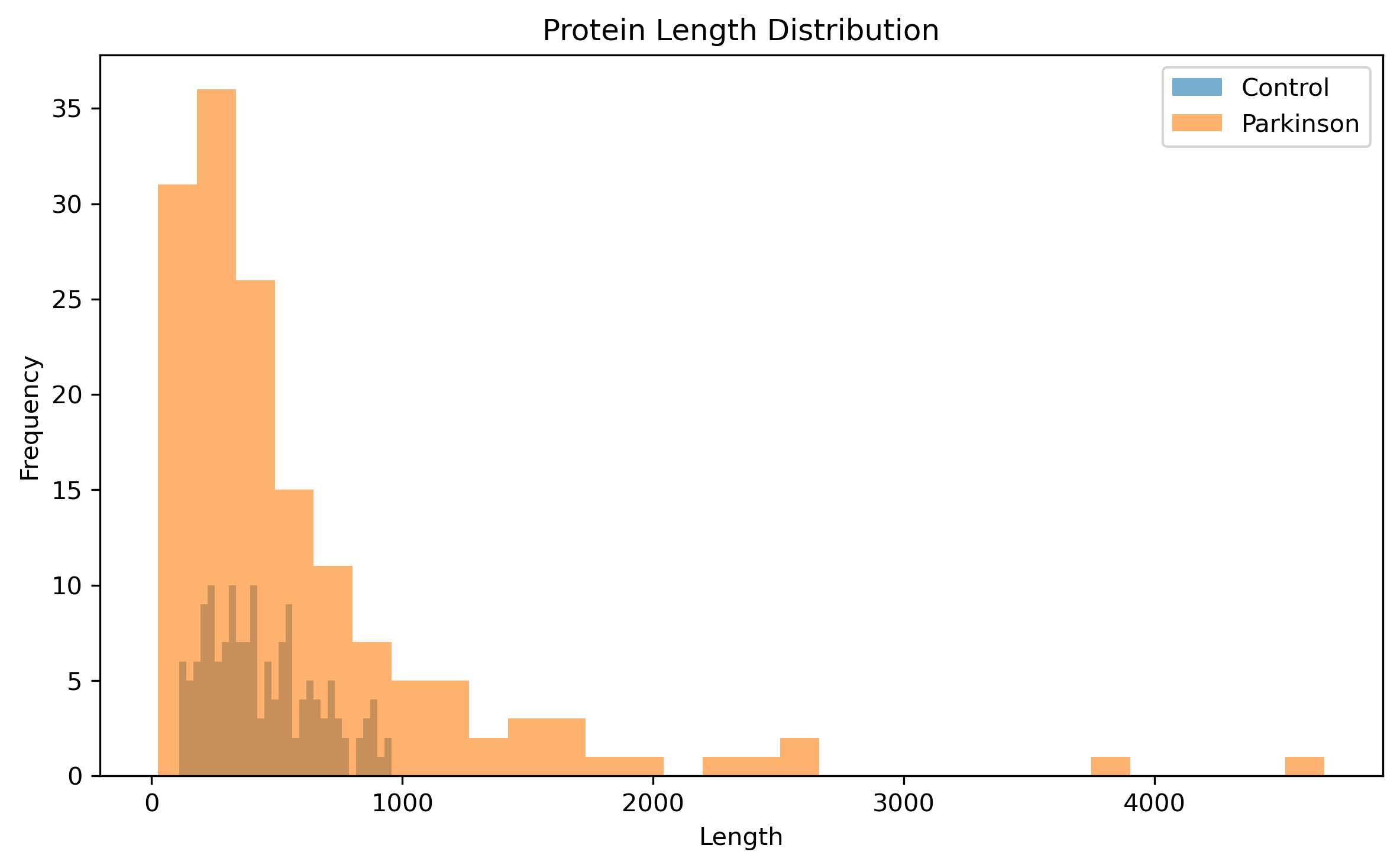}
	\caption{Histogram of protein length by class.}
	\label{fig:hist_longitud}
\end{figure}

Figure~\ref{fig:hist_longitud} further illustrates the distribution of sequence lengths, emphasizing the heavy-tailed behavior observed in the Parkinson class.

To further assess these differences, statistical tests were conducted. The Student's t-test indicates a statistically significant difference in means ($t = -2.872$, $p = 0.0044$). In contrast, the Mann--Whitney U test does not detect significant differences between distributions ($U = 11681.0$, $p = 0.8668$), suggesting that the observed difference in means is driven by distributional asymmetry and extreme values rather than a consistent shift across the entire distribution.

Overall, these results indicate that, although differences in mean length exist, protein length alone does not provide reliable class separability.

\subsubsection{Amino Acid Composition Analysis}

Amino acid composition was examined to assess differences in global residue distributions between classes. Each protein was represented by a 20-dimensional vector of relative amino acid frequencies.

\begin{figure}[H]
	\centering
	\includegraphics[width=0.7\linewidth]{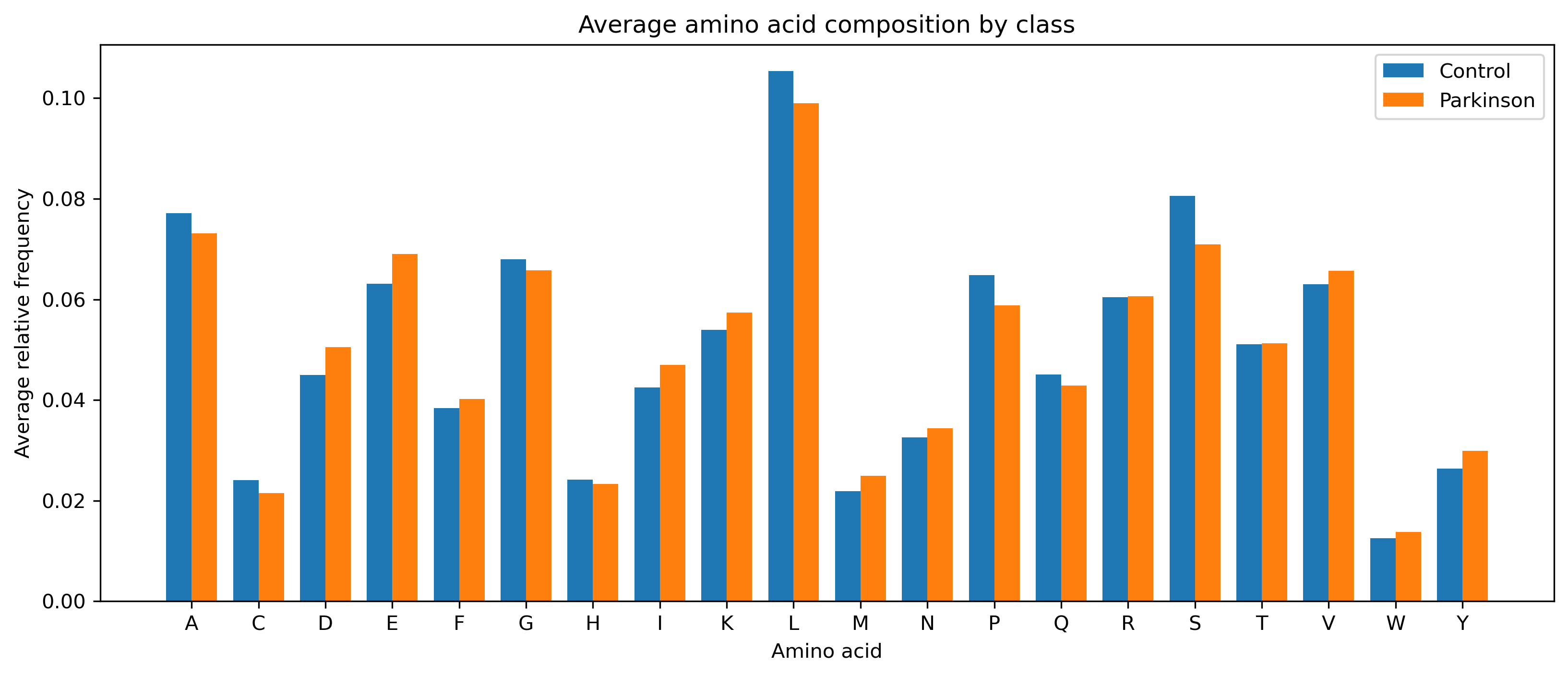}
	\caption{Average amino acid composition by class.}
	\label{fig:aa_composition}
\end{figure}

Figure~\ref{fig:aa_composition} compares the mean composition for both classes. Small but consistent variations are observed in several residues, including leucine (L), serine (S), glutamic acid (E), aspartic acid (D), alanine (A), glycine (G), and valine (V).

Despite these differences, their magnitude remains limited, and substantial overlap is observed in the distribution of individual amino acid frequencies across classes.

Overall, these results are consistent with the protein length analysis, indicating that global sequence descriptors capture variability in the data but do not provide meaningful class separability.

\subsection{Structural Analysis of the Feature Space}

The geometric structure of the feature space was analyzed to assess how different representations organize the data. In contrast to the previous exploratory analysis, which focuses on marginal distributions, this stage examines the spatial arrangement of samples and the presence of potential patterns in the representation space.

The dataset consists of 304 protein sequences evenly distributed between Parkinson-associated and control classes. Multiple representations were constructed with dimensionalities ranging from 20 features (amino acid composition) to 400 features (\textit{k}-mer descriptors). A reduced \textit{k}-mer representation obtained through genetic algorithm-based feature selection was also considered, resulting in a lower-dimensional space determined independently within each training fold.

Two complementary perspectives are considered:

\begin{itemize}
	\item \textbf{Geometric structure}, analyzed through dimensionality reduction techniques such as Principal Component Analysis (PCA), to examine the distribution of samples along directions of maximum variance.
	
	\item \textbf{Grouping behavior}, analyzed through unsupervised clustering methods, to evaluate how samples are partitioned in the feature space.
\end{itemize}

The following subsections present these analyses through PCA projections and clustering evaluations.

\subsubsection{Separability Analysis using PCA}

Principal Component Analysis (PCA) was applied to examine the structure of the feature representations in a reduced two-dimensional space.

\begin{figure}[H]
	\centering
	
	\begin{subfigure}{0.32\textwidth}
		\centering
		\includegraphics[width=\linewidth,height=4.5cm,keepaspectratio]{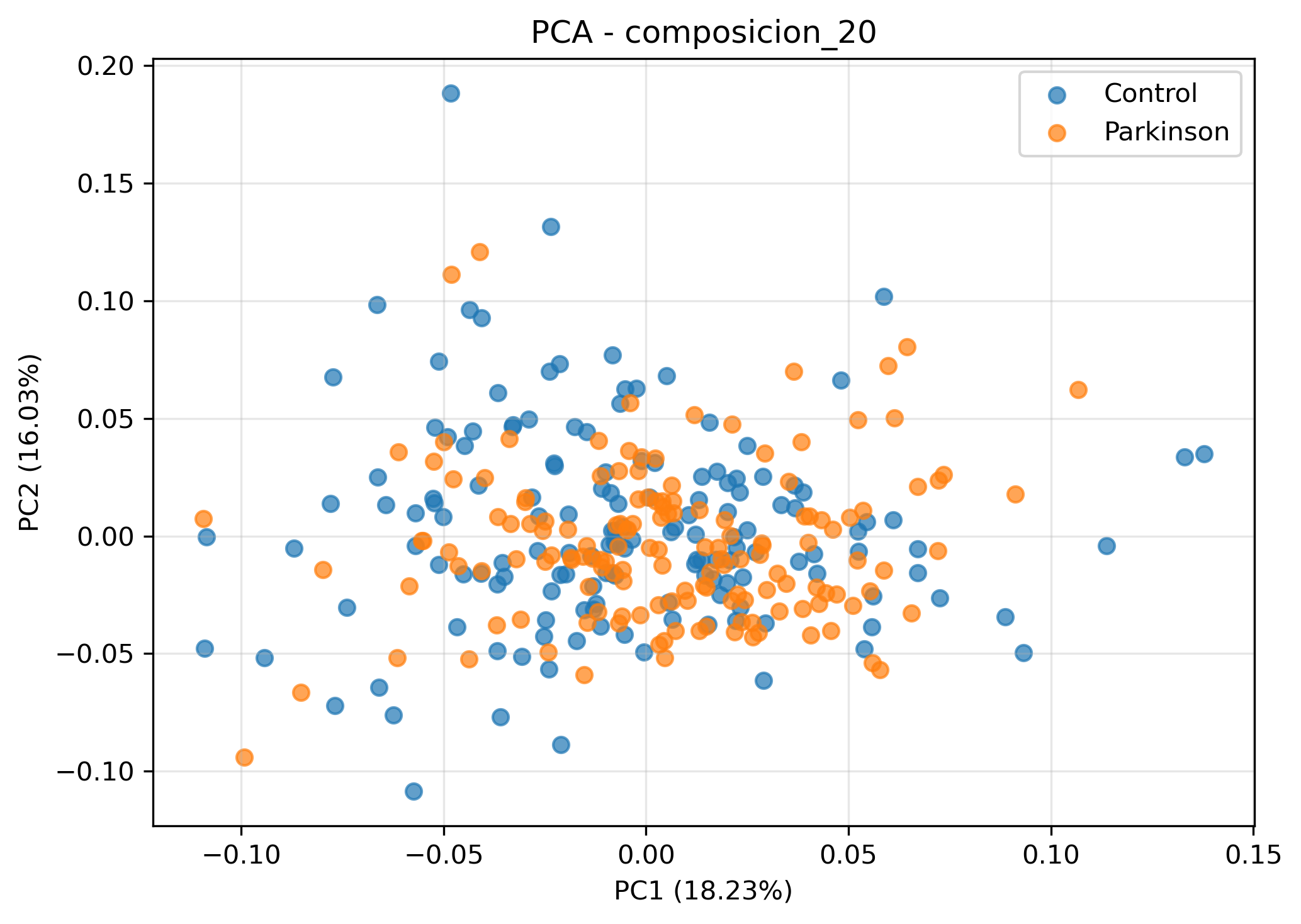}
		\caption{Amino acid composition}
	\end{subfigure}
	\hfill
	\begin{subfigure}{0.32\textwidth}
		\centering
		\includegraphics[width=\linewidth,height=4.5cm,keepaspectratio]{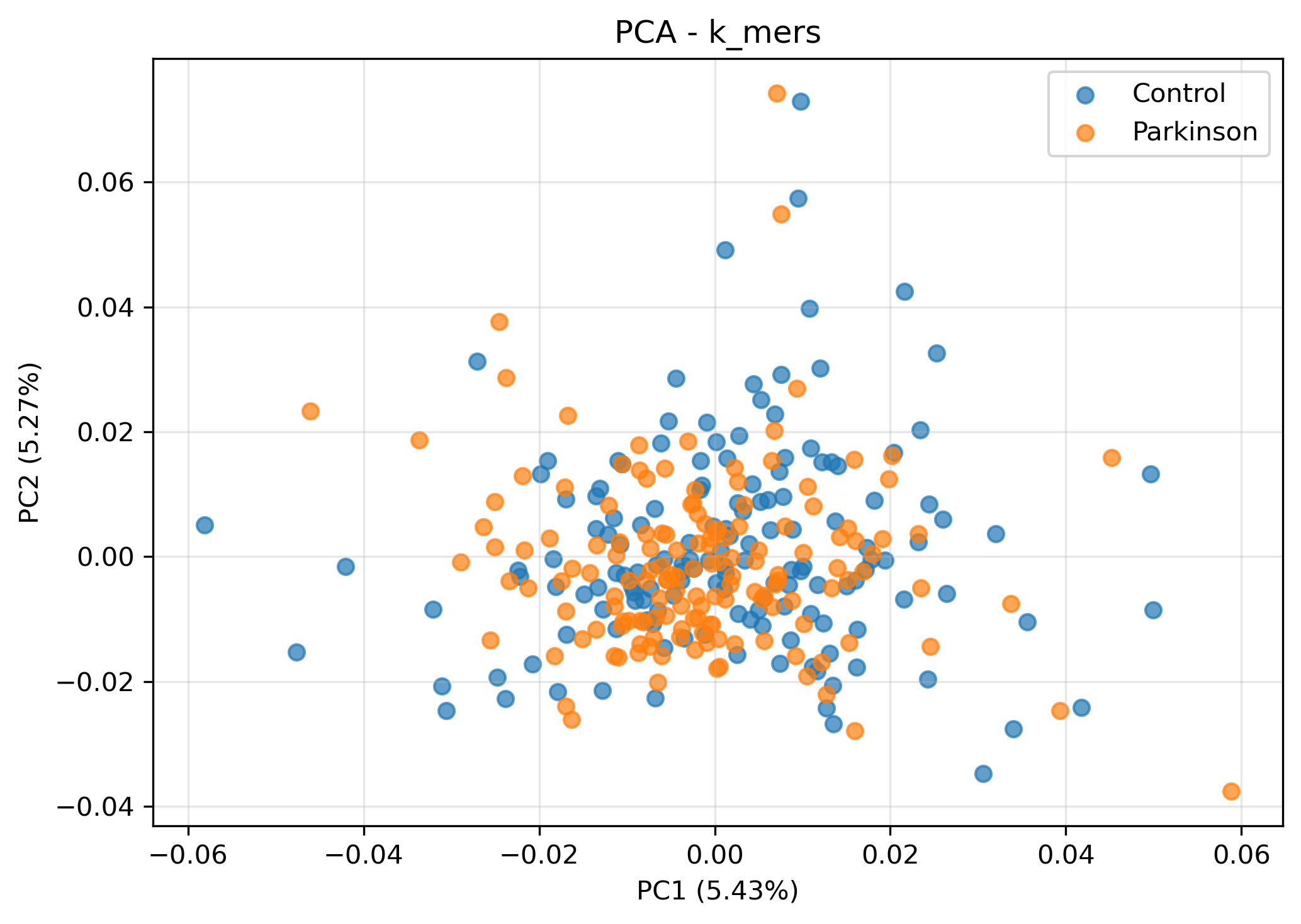}
		\caption{\textit{k}-mers ($k=2$)}
	\end{subfigure}
	\hfill
	\begin{subfigure}{0.32\textwidth}
		\centering
		\includegraphics[width=\linewidth,height=3.12cm,keepaspectratio]{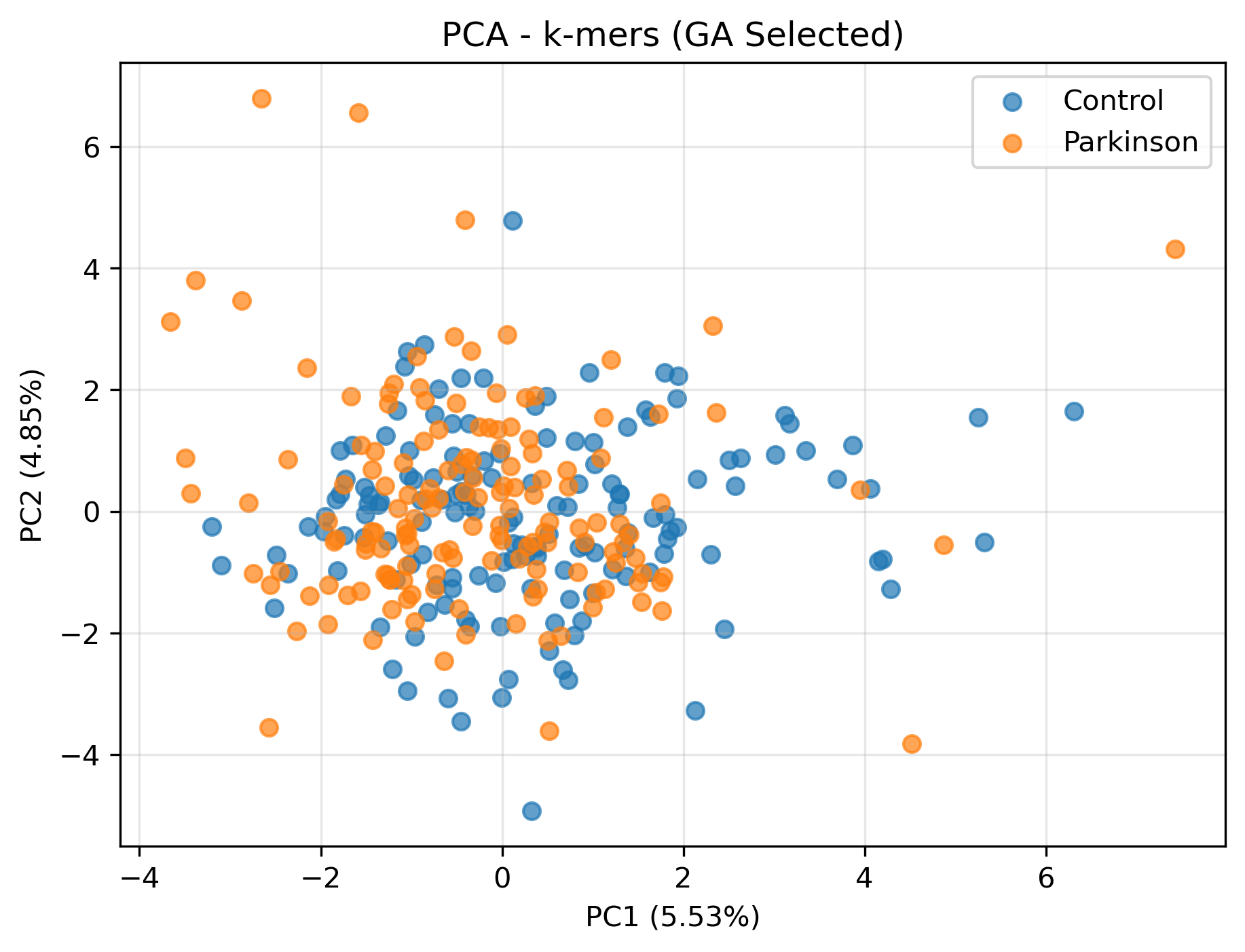}
		\caption{\textit{k}-mers with GA}
	\end{subfigure}
	
	\caption{PCA projections for different feature representations, showing substantial overlap between Parkinson-associated and control proteins across all cases.}
	\label{fig:pca_comparacion}
\end{figure}

Figure~\ref{fig:pca_comparacion} presents PCA projections for three representative cases: amino acid composition (20-dimensional), \textit{k}-mer-based descriptors ($k=2$, 400-dimensional), and \textit{k}-mers after feature selection using a genetic algorithm.

The first two principal components capture only a limited proportion of the total variance across all representations, indicating that the two-dimensional projection provides a partial view of the underlying structure. Nevertheless, no clear separation between classes is observed even along these directions of maximum variance.

Across all representations, the projected data exhibit strong overlap between classes, with no distinct boundaries or cluster-like structures. The distributions appear continuous and highly intermixed.

For amino acid composition, the projection forms a relatively compact distribution, consistent with the lower dimensionality of the representation, with both classes occupying largely overlapping regions.

The \textit{k}-mer representation shows greater dispersion, reflecting its higher dimensionality and the incorporation of more detailed sequence patterns. However, this increased variability does not translate into improved class separability.

After applying feature selection using a genetic algorithm, the dimensionality is reduced, but the overall structure remains largely unchanged, with no observable improvement in class separation.

\subsubsection{Clustering Analysis}

Clustering techniques were applied to evaluate how different representations organize the data without using class labels during training. Two methods were considered: K-Means and Agglomerative clustering.

Table~\ref{tab:clustering} summarizes the clustering performance across representations using both internal and external validation metrics.

\begin{table*}[t]
	\centering
	\caption{Clustering performance across representations.}
	\label{tab:clustering}
	\small
	\resizebox{\textwidth}{!}{
		\begin{tabular}{l l c c c c c c}
			\toprule
			\textbf{Representation} & \textbf{Method} & \textbf{Silhouette} & \textbf{ARI} & \textbf{NMI} & \textbf{Accuracy} & \textbf{F1} & \textbf{Specificity} \\
			\midrule
			k-mers & K-Means & 0.0173 & \textbf{0.0181} & 0.0185 & \textbf{0.5724} & \textbf{0.6389} & 0.3882 \\
			k-mers & Agglomerative & 0.4925 & 0.0000 & 0.0064 & 0.5033 & 0.0131 & \textbf{1.0000} \\
			Amino acid composition & K-Means & 0.1218 & 0.0143 & 0.0140 & 0.5658 & 0.6185 & 0.4276 \\
			Amino acid composition & Agglomerative & 0.1127 & 0.0146 & 0.0161 & 0.5658 & 0.6393 & 0.3618 \\
			Physicochemical properties & K-Means & 0.1743 & 0.0092 & 0.0090 & 0.5559 & 0.5603 & 0.5461 \\
			Physicochemical properties & Agglomerative & 0.1575 & 0.0031 & 0.0049 & 0.5395 & 0.5882 & 0.4211 \\
			Length (original + log) & K-Means & 0.5389 & 0.0046 & 0.0068 & 0.5428 & 0.4232 & 0.7500 \\
			Length (original + log) & Agglomerative & \textbf{0.7045} & 0.0079 & \textbf{0.0751} & 0.5461 & 0.1687 & \textbf{1.0000} \\
			\bottomrule
		\end{tabular}
	}
\end{table*}

Internal validation was assessed using the Silhouette coefficient, which measures cluster compactness and separation independently of class labels. While some representations, such as length-based features, exhibit relatively high Silhouette values, this reflects internal geometric structure rather than meaningful alignment with the underlying classes.

External validation metrics, including Adjusted Rand Index (ARI) and Normalized Mutual Information (NMI), remain consistently close to zero across all configurations. This indicates that the clustering structure is largely unrelated to the ground-truth class labels.

Classification-oriented metrics derived from cluster assignments (accuracy, precision, recall, F1-score, and specificity) show unstable and inconsistent behavior across representations. In several cases, extreme values are observed (e.g., perfect specificity or very low F1-score), reflecting degenerate clustering solutions rather than meaningful class separation.

Overall, the results indicate that none of the evaluated representations induces a clustering structure consistent with the underlying class labels. These findings are consistent with the PCA analysis and suggest that the dominant sources of variation captured by sequence-derived features are not aligned with the classification objective.

\subsection{Supervised Classification Results}

This section reports the classification performance obtained under the staged evaluation protocol described in Section~\ref{sec:exp_design}. The experimental design follows a progressive, stage-based approach, in which models are selected and refined according to their performance in earlier stages, rather than exhaustively evaluating all possible model–representation combinations.

The results are organized into two groups: (i) base representations evaluated under both baseline cross-validation and nested cross-validation settings, and (ii) advanced and refined representations evaluated using targeted model configurations.

\subsubsection{Base Representations}

Table~\ref{tab:base_models} summarizes the performance of baseline models evaluated without hyperparameter optimization, while Table~\ref{tab:nested_models} presents the results obtained after applying nested cross-validation to identify optimized configurations for each representation.

\begin{table*}[t]
	\centering
	\small
	\renewcommand{\arraystretch}{1.3}
	\setlength{\tabcolsep}{4pt}
	\caption{Best-performing models for base representations (without hyperparameter optimization). Results are reported as mean $\pm$ standard deviation across folds.}
	\label{tab:base_models}
	
	\resizebox{\textwidth}{!}{
		\begin{tabular}{l l c c c c c c}
			\toprule
			\textbf{Representation} & \textbf{Model} & \textbf{Accuracy} & \textbf{Precision} & \textbf{Recall} & \textbf{F1-score} & \textbf{ROC-AUC} & \textbf{PR-AUC} \\
			\midrule
			
			\textit{k}-mers ($k=2$) & KNN & 0.5066 $\pm$ 0.0223 & 0.5036 $\pm$ 0.0147 & 0.9802 $\pm$ 0.0162 & \textbf{0.6652 $\pm$ 0.0144} & 0.5382 $\pm$ 0.0359 & 0.5215 $\pm$ 0.0258 \\
			
			Amino acid composition & KNN & 0.6050 $\pm$ 0.0451 & 0.5879 $\pm$ 0.0354 & 0.6899 $\pm$ 0.0772 & 0.6343 $\pm$ 0.0529 & 0.6057 $\pm$ 0.0363 & 0.5801 $\pm$ 0.0312 \\
			
			Length (original + log) & MLP (shallow) & 0.6349 $\pm$ 0.0543 & 0.6726 $\pm$ 0.0933 & 0.5527 $\pm$ 0.0240 & 0.6041 $\pm$ 0.0407 & 0.6561 $\pm$ 0.0637 & 0.7186 $\pm$ 0.0470 \\
			
			Physicochemical properties & Logistic Regression & 0.5954 $\pm$ 0.0581 & 0.5941 $\pm$ 0.0548 & 0.6049 $\pm$ 0.0701 & 0.5985 $\pm$ 0.0584 & 0.6338 $\pm$ 0.0564 & 0.6706 $\pm$ 0.0429 \\
			
			\bottomrule
		\end{tabular}
	}
\end{table*}

\begin{table*}[t]
	\centering
	\small
	\renewcommand{\arraystretch}{1.3}
	\setlength{\tabcolsep}{4pt}
	\caption{Best-performing models for base representations under nested cross-validation. Results are reported as mean $\pm$ standard deviation across folds.}
	\label{tab:nested_models}
	
	\resizebox{\textwidth}{!}{
		\begin{tabular}{l l c c c c c c}
			\toprule
			\textbf{Representation} & \textbf{Model} & \textbf{Accuracy} & \textbf{Precision} & \textbf{Recall} & \textbf{F1-score} & \textbf{ROC-AUC} & \textbf{PR-AUC} \\
			\midrule
			
			\textit{k}-mers ($k=2$) & KNN & 0.5099 $\pm$ 0.0330 & 0.5053 $\pm$ 0.0211 & 0.9800 $\pm$ 0.0183 & \textbf{0.6667 $\pm$ 0.0222} & 0.5341 $\pm$ 0.0449 & 0.5203 $\pm$ 0.0299 \\
			
			Amino acid composition & KNN & 0.5755 $\pm$ 0.0471 & 0.5597 $\pm$ 0.0381 & 0.6897 $\pm$ 0.0932 & 0.6170 $\pm$ 0.0590 & 0.6085 $\pm$ 0.0669 & 0.5991 $\pm$ 0.0595 \\
			
			Physicochemical properties & SVM & 0.5493 $\pm$ 0.0902 & 0.5523 $\pm$ 0.0863 & 0.7249 $\pm$ 0.2490 & 0.6057 $\pm$ 0.0980 & 0.5120 $\pm$ 0.1869 & 0.5711 $\pm$ 0.1397 \\
			
			Length (original + log) & Logistic Regression & 0.6283 $\pm$ 0.0633 & 0.6888 $\pm$ 0.1053 & 0.4800 $\pm$ 0.0462 & 0.5647 $\pm$ 0.0640 & 0.6603 $\pm$ 0.0563 & 0.7199 $\pm$ 0.0494 \\
			
			\bottomrule
		\end{tabular}
	}
\end{table*}

Across base representations, performance remains limited and relatively stable across both evaluation settings. The comparison between Tables~\ref{tab:base_models} and~\ref{tab:nested_models} shows that hyperparameter optimization leads to only minor variations in performance, with no consistent improvement across representations.

The \textit{k}-mer representation achieves the highest F1-score in both settings. However, this result is driven by a strong imbalance between precision and recall, characterized by very high recall and low precision. This behavior indicates a strong bias toward the positive class rather than effective discriminative capability.

This effect is further reflected in the confusion matrix analysis. The \textit{k}-mer representation exhibits extremely high sensitivity (approximately 0.98) but very low specificity (approximately 0.03), indicating that most samples are classified as positive, resulting in a large number of false positives.

In contrast, the length-based representation shows more balanced behavior, with lower sensitivity (approximately 0.55) and higher specificity (approximately 0.71), suggesting improved discrimination of negative instances. Amino acid composition and physicochemical representations exhibit intermediate performance, with moderate sensitivity and specificity.

Overall, differences between representations remain moderate, and none of the evaluated base representations achieves strong discriminative performance. These results indicate that global and low-level sequence descriptors are insufficient to reliably separate the two classes.

\subsubsection{Advanced and Refined Representations}

Table~\ref{tab:advanced_models} summarizes the classification performance obtained for advanced and refined representations under nested cross-validation.

\begin{table*}[t]
	\centering
	\small
	\renewcommand{\arraystretch}{1.3}
	\setlength{\tabcolsep}{4pt}
	\caption{Best-performing models for advanced and refined representations under nested cross-validation. Each row reports the best model identified for the corresponding representation. Results are reported as mean $\pm$ standard deviation across folds.}
	\label{tab:advanced_models}
	
	\resizebox{\textwidth}{!}{
		\begin{tabular}{l l c c c c c c}
			\toprule
			\textbf{Representation} & \textbf{Model} & \textbf{Accuracy} & \textbf{Precision} & \textbf{Recall} & \textbf{F1-score} & \textbf{ROC-AUC} & \textbf{PR-AUC} \\
			\midrule
			
			5 most frequent amino acids & SVM (RBF) & 0.5460 $\pm$ 0.0699 & 0.5535 $\pm$ 0.0860 & 0.6923 $\pm$ 0.1790 & 0.6002 $\pm$ 0.0633 & 0.5330 $\pm$ 0.1399 & 0.5793 $\pm$ 0.0985 \\
			
			Hybrid representation & KNN & 0.5263 $\pm$ 0.0369 & 0.5143 $\pm$ 0.0242 & 0.9469 $\pm$ 0.0382 & 0.6665 $\pm$ 0.0283 & 0.5796 $\pm$ 0.0677 & 0.5643 $\pm$ 0.0520 \\
			
			\textit{k}-mers + GA & KNN & 0.5001 $\pm$ 0.0844 & 0.4976 $\pm$ 0.0603 & 0.8209 $\pm$ 0.1598 & 0.6177 $\pm$ 0.0873 & 0.5403 $\pm$ 0.0857 & 0.5438 $\pm$ 0.0674 \\
			
			ProtBERT embeddings & KNN & 0.6646 $\pm$ 0.0282 & 0.6446 $\pm$ 0.0276 & 0.7366 $\pm$ 0.0417 & 0.6870 $\pm$ 0.0254 & 0.7383 $\pm$ 0.0281 & 0.7329 $\pm$ 0.0609 \\
			
			ProtBERT embeddings & MLP (shallow) & 0.7041 $\pm$ 0.0313 & 0.7048 $\pm$ 0.0386 & 0.7041 $\pm$ 0.0202 & \textbf{0.7043 $\pm$ 0.0284} & 0.7480 $\pm$ 0.0474 & 0.7573 $\pm$ 0.0262 \\
			
			\bottomrule
		\end{tabular}
	}
\end{table*}

Across the evaluated advanced representations, performance shows moderate variation, with F1-scores ranging from 0.6002 to 0.7043. The lowest performance is observed for the reduced representation based on the five most frequent amino acids, while higher values are obtained for representations incorporating richer sequence information.

The hybrid representation and the \textit{k}-mer representation with genetic algorithm selection achieve intermediate performance levels (F1-scores of 0.6665 and 0.6177, respectively). However, in both cases, these results are driven by high recall values (0.9469 and 0.8209) combined with lower precision, indicating a persistent bias toward positive predictions rather than effective class discrimination.

This behavior is further reflected in sensitivity and specificity. The hybrid representation exhibits very high sensitivity but low specificity (approximately 0.11), indicating a high rate of false positives. A similar pattern is observed for the \textit{k}-mer representation with genetic algorithm selection, where specificity remains limited (approximately 0.18), despite moderate improvements over the full \textit{k}-mer representation.

The reduced representation based on the five most frequent amino acids shows slightly more balanced behavior, with moderate specificity (approximately 0.40), although overall discriminative performance remains limited.

In contrast, ProtBERT-based models exhibit a substantially improved balance between sensitivity and specificity. The KNN model achieves a specificity of approximately 0.59, while the MLP model reaches approximately 0.70, indicating a more stable classification behavior.

The highest performance is obtained using ProtBERT embeddings. The KNN model achieves an F1-score of 0.6870, while the MLP model further improves this result to 0.7043. Unlike other representations, ProtBERT-based models maintain a more balanced relationship between precision and recall.

Additional performance indicators follow a consistent trend. Considering the best-F1 configuration selected for each representation, ROC-AUC values range from 0.5330 to 0.7480, while PR-AUC values range from 0.5793 to 0.7573, further supporting the improved discriminative capacity of contextual embeddings.

\begin{figure}[t]
	\centering
	
	\begin{subfigure}{0.40\textwidth}
		\centering
		\includegraphics[width=\linewidth]{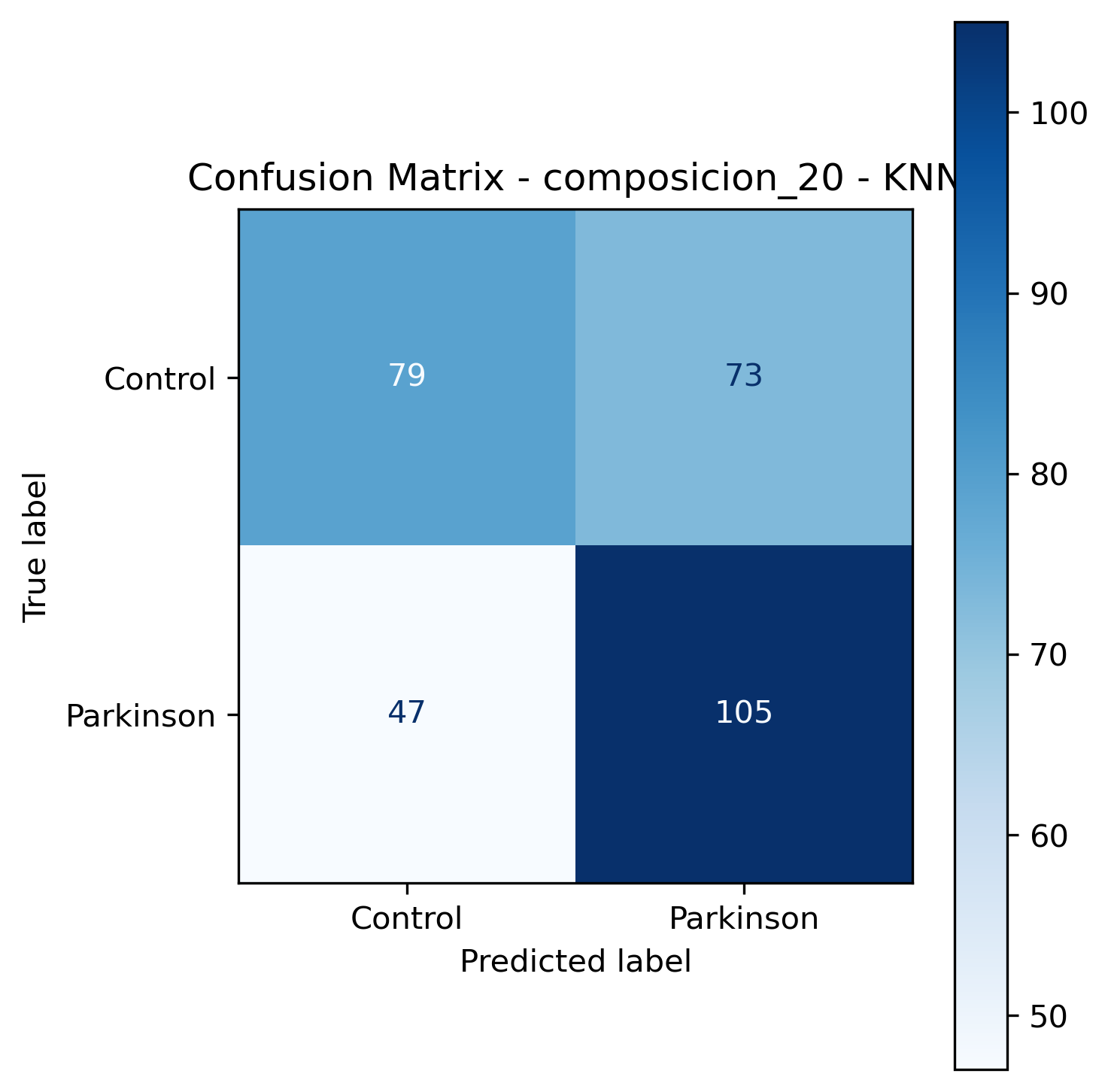}
		\caption{KNN with amino acid composition}
	\end{subfigure}
	\hfill
	\begin{subfigure}{0.40\textwidth}
		\centering
		\includegraphics[width=\linewidth]{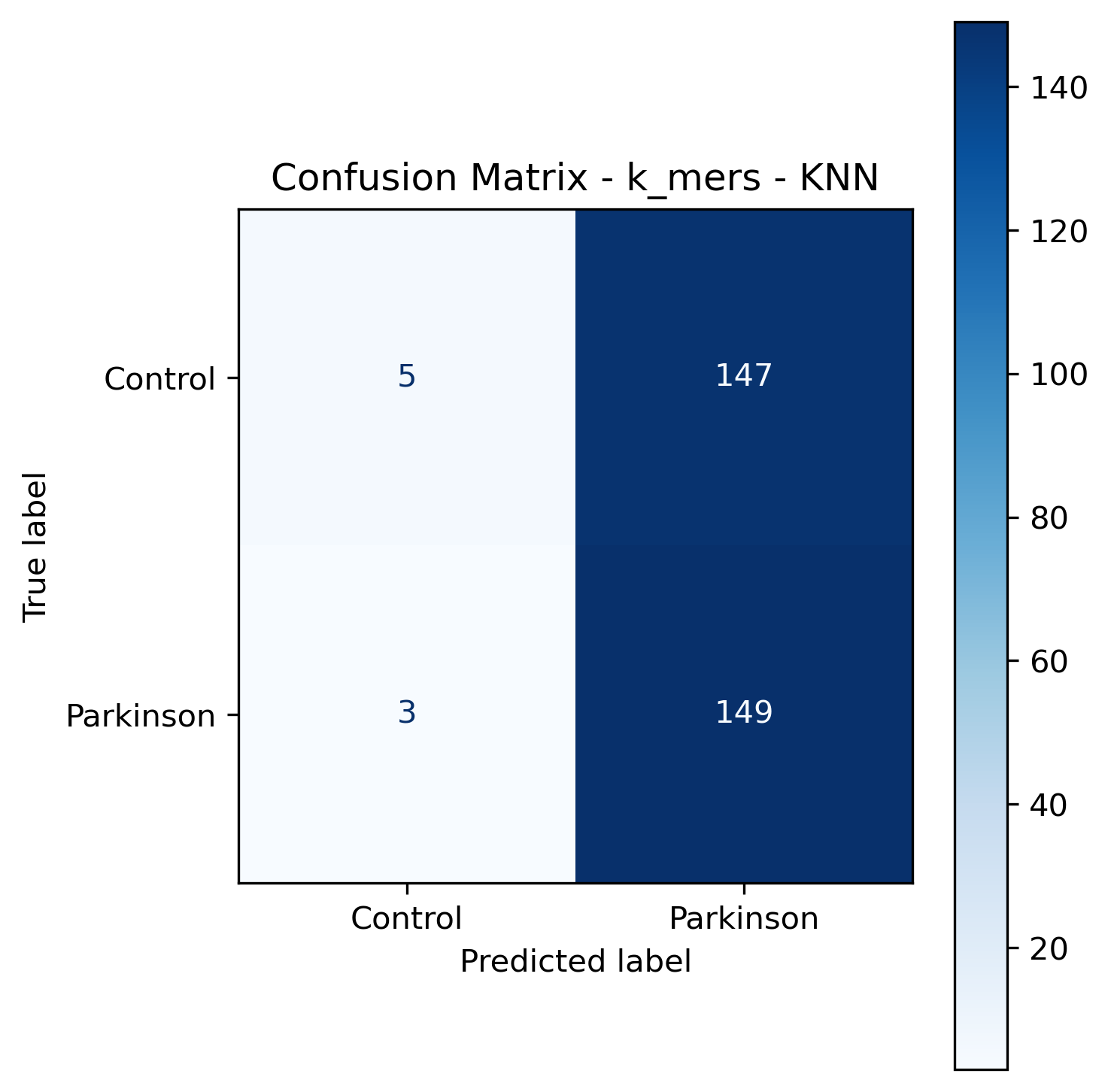}
		\caption{KNN with \textit{k}-mers}
	\end{subfigure}
	
	\vspace{0.2cm}
	
	\begin{subfigure}{0.40\textwidth}
		\centering
		\includegraphics[width=\linewidth]{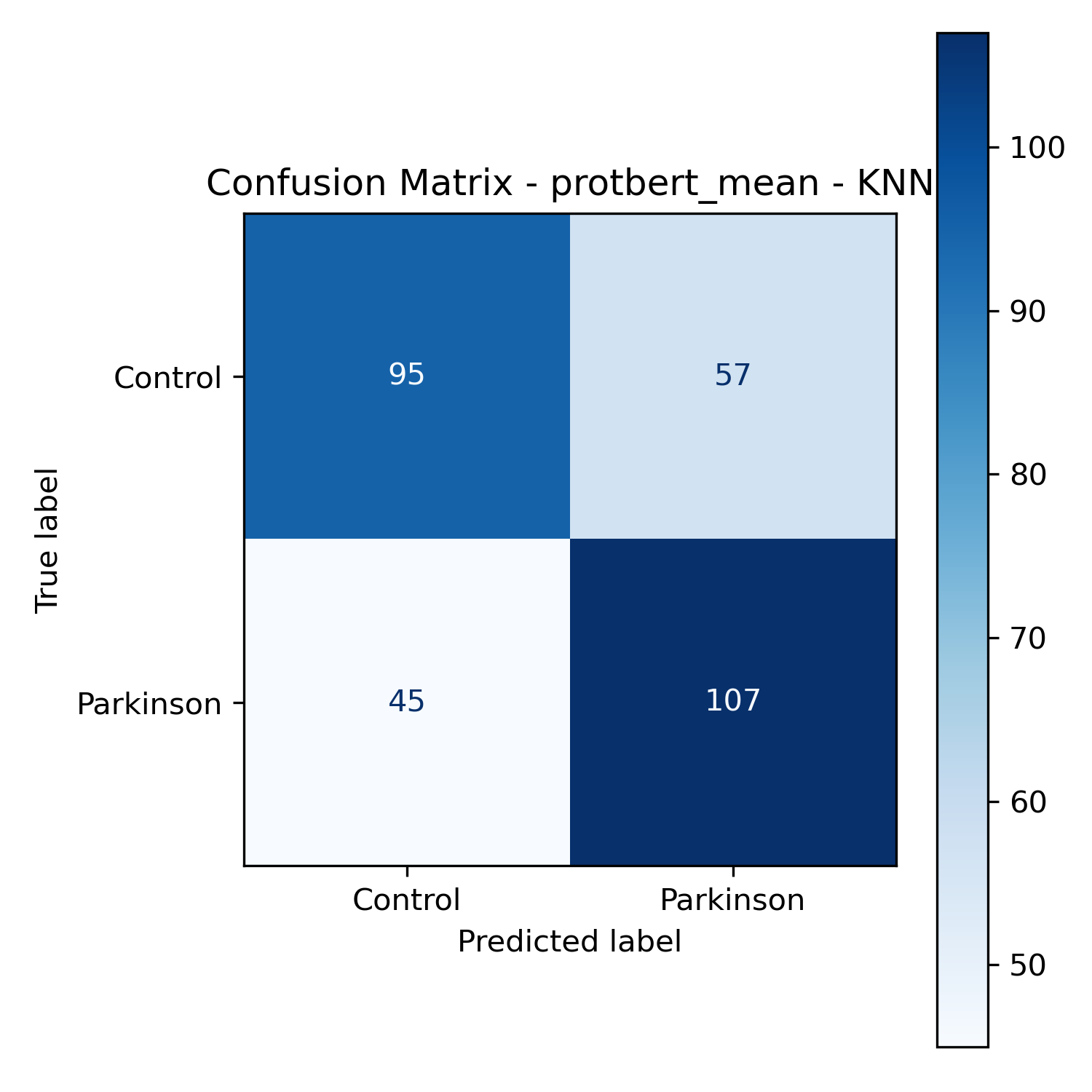}
		\caption{KNN with ProtBERT embeddings}
	\end{subfigure}
	\hfill
	\begin{subfigure}{0.40\textwidth}
		\centering
		\includegraphics[width=\linewidth]{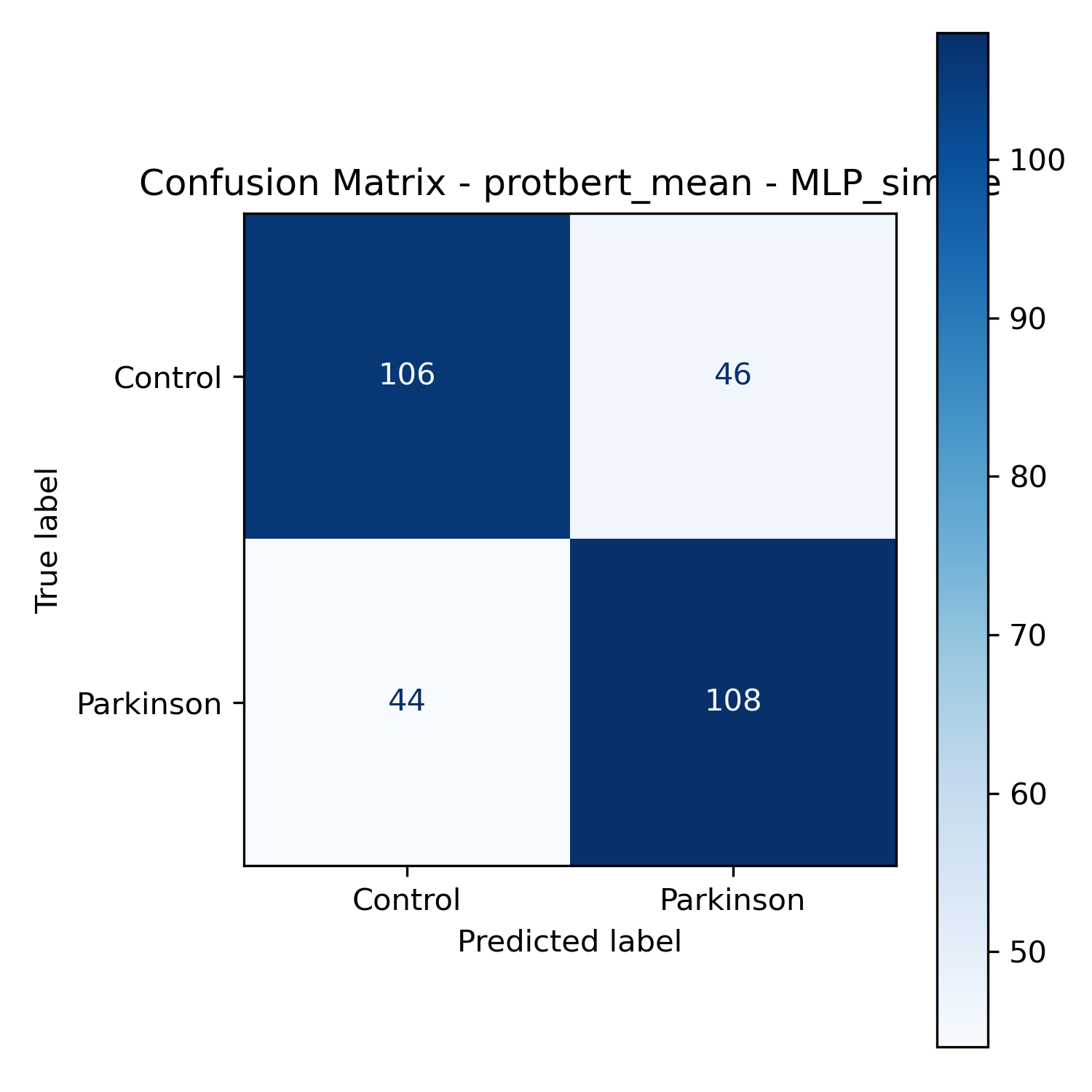}
		\caption{MLP with ProtBERT embeddings}
	\end{subfigure}
	
	\caption{Confusion matrices for representative models across different feature representations.}
	\label{fig:cm_comparison}
\end{figure}

To further examine classification behavior, Figure~\ref{fig:cm_comparison} presents confusion matrices for representative models across different feature representations, including the best-performing ProtBERT-based MLP configuration.

Overall, while advanced representations such as ProtBERT embeddings provide measurable improvements in performance and class balance, the achieved results remain moderate. These findings indicate that, although contextual embeddings capture richer sequence information, primary sequence data alone remains insufficient to achieve strong discriminative performance in this classification task.

\subsubsection{Detailed Model Behavior: KNN Analysis}

To further analyze class-level behavior, the KNN model using the amino acid composition representation was examined in detail.

\begin{figure}[H]
	\centering
	\includegraphics[width=0.4\linewidth]{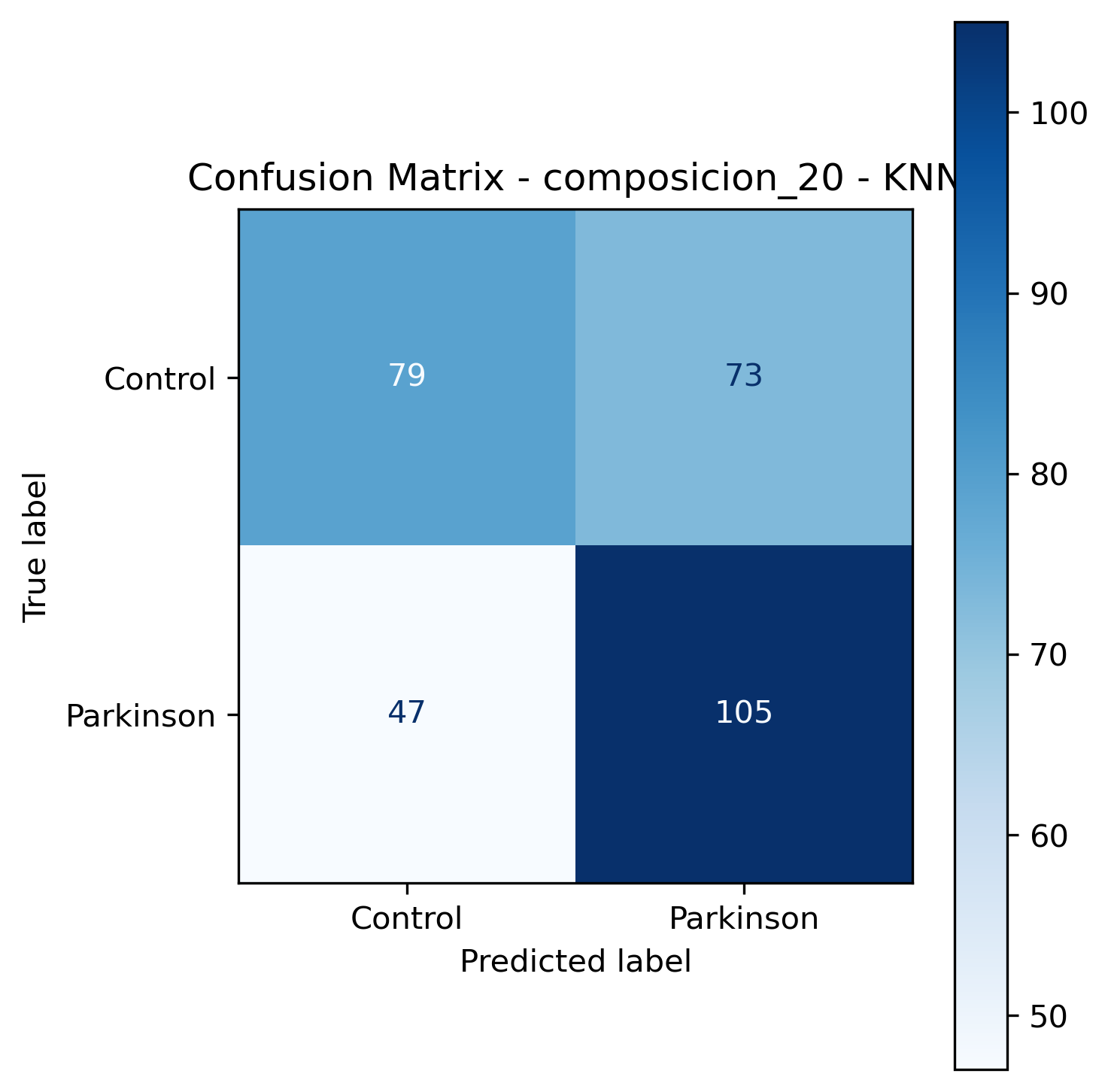}
	\caption{Confusion matrix for KNN with amino acid composition.}
	\label{fig:cm_knn_opt}
\end{figure}

Figure~\ref{fig:cm_knn_opt} shows the confusion matrix corresponding to the best-performing configuration of this model.

The confusion matrix reveals that the model correctly identifies a substantial proportion of Parkinson-associated proteins, as reflected by the number of true positives. However, a considerable number of control proteins are misclassified as positive, leading to an elevated number of false positives.

This pattern indicates an asymmetric classification behavior. The model favors the positive class, achieving high sensitivity while exhibiting reduced precision and moderate specificity. As a result, the model is effective at detecting positive instances but struggles to correctly identify negative samples.

Overall, these results highlight a systematic bias toward positive predictions and reinforce the limitations of global compositional features for achieving balanced and reliable class discrimination.

\subsubsection{Statistical Comparison of Models}

To assess differences in performance across models and representations, the F1-score values obtained across cross-validation folds were compared using non-parametric statistical testing.

The results show that performance differences between models remain relatively small across representations. Most configurations yield F1-scores within a narrow range, approximately from 0.6002 to 0.7043, with moderate variability across folds.

To formally evaluate these differences, the Friedman test was applied across models and representations. The test did not reveal statistically significant differences ($p = 0.1749$), indicating that the observed performance variations are not sufficient to support the superiority of any particular model or representation.

These findings are consistent with the overall similarity observed in performance metrics, suggesting that improvements across models and feature representations are generally incremental rather than substantial.

\subsubsection{Error Analysis}

Error analysis was performed to examine misclassification patterns across models and representations.

Misclassified instances were analyzed through confusion matrices and prediction outputs, with particular attention to false positives and false negatives.

Across models, a consistent pattern is the prevalence of false positives, where control proteins are classified as Parkinson-associated. This behavior is especially pronounced in representations such as \textit{k}-mers and hybrid features, which achieve high recall values (up to approximately 0.98 and 0.95, respectively) but low specificity, indicating a systematic tendency to over-predict the positive class.

False negatives are also present, although their frequency varies depending on the representation. Models with more balanced recall values, such as those based on length features or ProtBERT embeddings, exhibit a more even distribution of errors between classes.

Overall, error patterns remain relatively consistent across representations, with no configuration achieving clear class separation. The observed F1-scores, ranging from approximately 0.60 to 0.70, reflect the persistence of misclassification across all evaluated models.

These results suggest that classification errors are not driven by isolated model deficiencies but rather by the limited discriminative signal available in sequence-based representations.

\begin{figure}[H]
	\centering
	\includegraphics[width=0.7\linewidth]{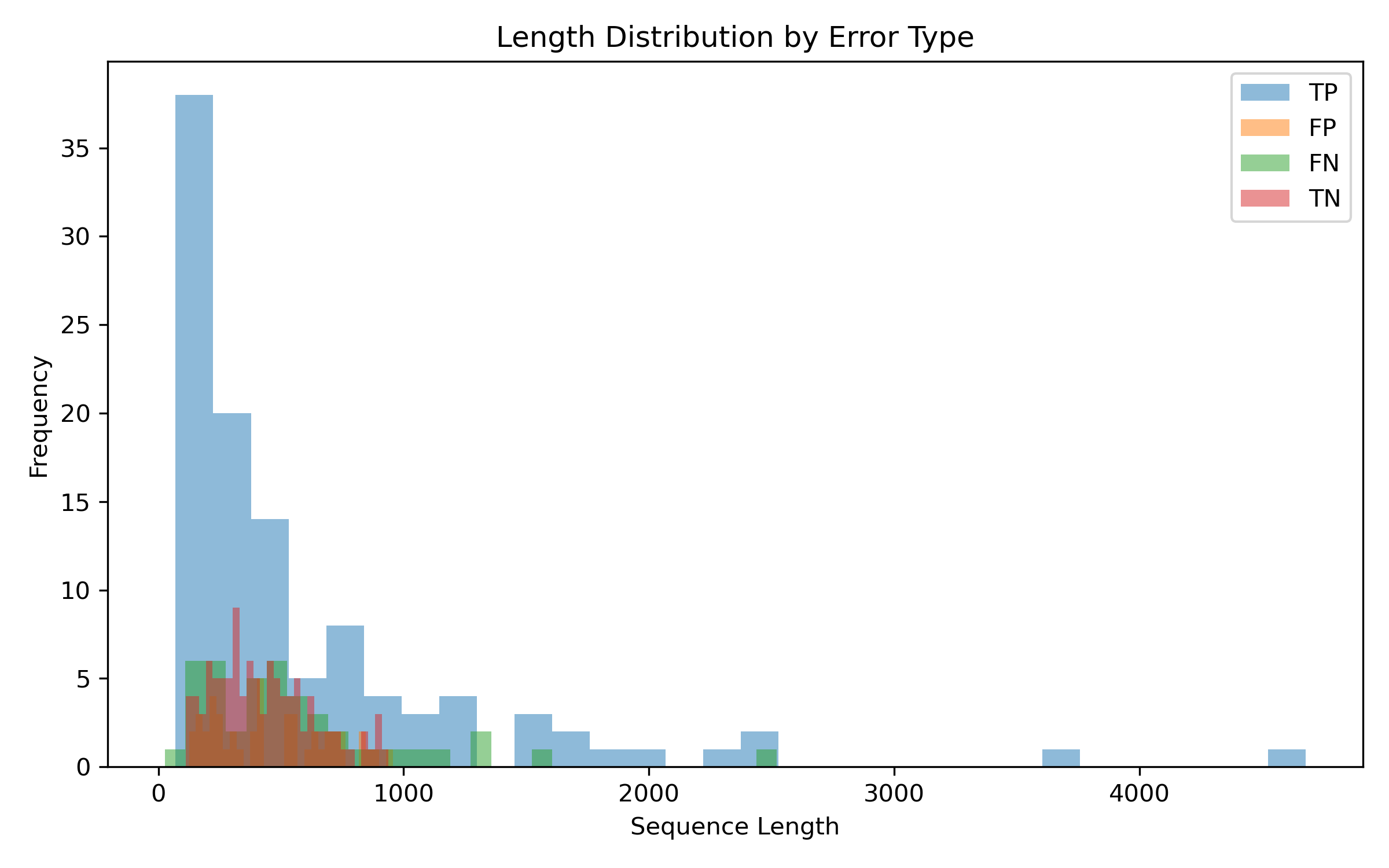}
	\caption{Distribution of sequence length by prediction type (TP, FP, FN, TN) for the ProtBERT + MLP configuration.}
	\label{fig:error_length_distribution}
\end{figure}

To further examine the relationship between sequence properties and classification outcomes, Figure~\ref{fig:error_length_distribution} presents the distribution of protein length for different prediction types (true positives, false positives, false negatives, and true negatives) in the ProtBERT + MLP configuration.

The distributions show substantial overlap between prediction types, indicating that sequence length alone does not explain the observed classification errors. This further supports the conclusion that global sequence descriptors provide limited discriminatory information for this task.

\section{Discussion}
\label{sec:discusion}

The results obtained in this study provide a consistent and robust characterization of the limitations of sequence-derived representations for distinguishing Parkinson's disease-associated proteins from control proteins. Across all experiments, classification performance remains constrained, indicating that primary sequence information alone is insufficient to achieve strong discriminative capability in this task.

\paragraph{Intrinsic nature of the problem}

A central finding is that classification performance is primarily limited by the information content of the representations rather than by model capacity. Although more expressive models, such as MLP applied to ProtBERT embeddings, achieve the highest performance, the observed improvements remain moderate. This is supported by the observation that performance values across models and representations remain within a narrow range (F1 $\approx$ 0.60--0.70), even when using high-capacity models. This suggests that increasing model complexity does not fundamentally alter the separability of the problem, indicating that the discriminative signal available in the input space is inherently weak.

\paragraph{Structure of the feature space}

The lack of class separability is reflected in the geometric organization of the feature space. Across all representations, including high-dimensional embeddings, samples from both classes remain highly intermixed. Dimensionality reduction through PCA produces overlapping projections, and clustering results show minimal agreement with ground-truth labels, as evidenced by ARI and NMI values close to zero. These findings indicate that the dominant axes of variation are not aligned with the classification objective, reinforcing the absence of a well-defined class structure.

\paragraph{Effect of representation and dimensionality}

Increasing representational richness leads to only partial improvements. While \textit{k}-mer-based and hybrid representations capture additional local and global patterns, their performance is often associated with asymmetric classification behavior, characterized by high recall and reduced specificity. This indicates that additional features tend to amplify class-dependent biases (e.g., favoring positive predictions) without improving true discriminative separability.

Similarly, genetic algorithm-based feature selection reduces redundancy in the \textit{k}-mer space but does not yield consistent performance gains. The variability in selected subsets across folds further suggests that no stable subset of features captures robust discriminative information, highlighting the absence of representation-invariant signals.

\paragraph{Model behavior}

The observed model behavior reflects these representational limitations. Distance-based methods such as KNN are particularly sensitive to local overlap in the feature space, often leading to biased predictions toward the positive class. In contrast, nonlinear models such as MLP exhibit more balanced behavior, especially when applied to contextual embeddings. However, these improvements remain constrained by the same underlying data structure. The persistence of classification errors across models suggests that refinements in decision boundaries are insufficient when class distributions are strongly overlapping.

\paragraph{Consistency across analytical perspectives}

A key strength of this study is the consistency of findings across multiple analytical perspectives. Exploratory data analysis, feature space visualization, clustering, supervised evaluation, and error analysis all converge toward the same conclusion: class differences are subtle and not strongly expressed in sequence-derived representations. This convergence strengthens the validity of the conclusions and suggests that the observed limitations are intrinsic to the data rather than artifacts of specific modeling choices.

\paragraph{Biological implications}

From a biological perspective, these results suggest that the determinants of Parkinson's disease association are not fully encoded at the level of the primary sequence. Instead, relevant discriminative signals are likely associated with higher levels of biological organization, such as protein tertiary structure, interaction networks, or cellular context. This observation is consistent with the multifactorial nature of Parkinson's disease, where molecular mechanisms such as protein aggregation, mitochondrial dysfunction, and complex cellular processes play a central role. These processes are not explicitly captured by sequence composition alone, which limits the effectiveness of sequence-based representations for this task.

\paragraph{Limitations}

This study presents several limitations. Although strict class balance and a nested cross-validation framework were adopted to obtain robust and unbiased estimates of generalization performance, the dataset size remains relatively moderate and may affect the stability of performance estimates as well as the reliability of statistical comparisons. Additionally, labels are derived from curated biological databases and may include indirect disease associations or annotation noise. Finally, the analysis is restricted to sequence-derived features, limiting the scope of the conclusions to this representation paradigm. These considerations should be taken into account when interpreting the results, particularly regarding the generalizability of the findings.

\paragraph{Summary}

Overall, the results indicate that improvements in model complexity and feature engineering yield only incremental gains, while the fundamental separability of the data remains largely unchanged across all evaluated conditions. These findings highlight the need to incorporate richer sources of biological information when addressing complex classification tasks, particularly those involving disease-related protein characterization.

\section{Conclusions}
\label{sec:conclusions}

This work presented a controlled and leakage-free benchmark for evaluating the discriminative capacity of representations derived exclusively from protein primary sequences for the classification of proteins associated with Parkinson's disease. By employing a nested cross-validation framework, the study provides reliable estimates of generalization performance while isolating the contribution of sequence-derived features.

The results indicate that primary sequence information alone is insufficient to achieve robust class discrimination in this setting. This limitation was consistently observed across all evaluated representations, including global descriptors, \textit{k}-mer-based features, and protein language model embeddings. Even the best-performing configuration (ProtBERT + MLP) achieved only moderate performance (F1 = 0.704, ROC-AUC = 0.748), without clear separation between classes.

These findings were supported by multiple complementary analyses, including feature space projections, clustering behavior, and supervised evaluation, all of which revealed substantial overlap between classes. Together, these results suggest that the observed limitations are intrinsic to the available sequence information rather than artifacts of specific modeling choices.

Overall, this study provides empirical evidence that discriminative signals associated with Parkinson's disease are not fully captured at the level of protein primary sequence alone. More robust disease modeling will likely require richer biological information, such as structural, functional, interaction-level, or cellular-context descriptors. The proposed benchmark establishes a reproducible baseline for future research in protein-based disease classification.

\section{Future Work}
\label{sec:future}

The findings of this study suggest several research directions for overcoming the limitations identified in sequence-derived representations.

A primary direction is the integration of additional sources of biological information into the modeling pipeline. Given the limited separability observed across all evaluated representations, incorporating structural descriptors (e.g., secondary and tertiary structure), functional annotations, and protein--protein interaction networks may provide complementary signals not captured at the primary sequence level. The development of multimodal architectures capable of integrating these heterogeneous data sources represents a natural extension of this work.

Another relevant direction is the task-specific adaptation of protein language models. In this study, ProtBERT embeddings were used without fine-tuning, which may limit their ability to capture task-relevant patterns. Future work should explore fine-tuning strategies under strict validation protocols, as well as the evaluation of alternative architectures such as ESM-based models, while ensuring that information leakage is properly controlled.

The incorporation of evolutionary information also constitutes a promising line of research. Features derived from multiple sequence alignments, conservation scores, or protein family profiles may encode functional constraints not explicitly represented in individual sequences, potentially enhancing discriminative capacity.

From a representation perspective, future work may investigate richer local sequence patterns through higher-order \textit{k}-mers (e.g., $k \geq 3$), combined with principled feature selection or dimensionality reduction techniques to mitigate redundancy and instability in high-dimensional spaces.

Finally, improving interpretability remains an open challenge. Identifying the contribution of specific sequence regions or learned features to model predictions could provide deeper insight into the relationship between sequence patterns and biological mechanisms, helping bridge the gap between computational predictions and biological interpretation.

Overall, future research should focus on integrating complementary biological information and developing representations that move beyond primary sequence features, which appear insufficient to capture the complexity of disease-related signals observed in this study.

\section{Declaration of generative AI and AI-assisted technologies in the manuscript preparation process}

During the preparation of this work, the authors used ChatGPT (Open-AI) to assist in improving the clarity and readability of the manuscript, including language refinement and organization of the text. After using this tool, the authors reviewed and edited the content as needed and take full responsibility for the content of the published article.

\bibliographystyle{elsarticle-num}
\bibliography{references}

\end{document}